\documentclass[manuscript]{aastex}
\usepackage{graphicx}
\usepackage{float}
\usepackage{color}
\usepackage[normalem]{ulem}
\usepackage{natbib} 
\usepackage{bm}

\def\hmi{Helioseismic and Magnetic Imager}
\def\aia{Atmospheric Imaging Assembly}

\shorttitle{}
\shortauthors{Zhao et al.}

\begin{document}

\title{Observational Evidence of Magnetic Reconnection for {\bf Brightenings and Transition Region Arcades }in IRIS observations}

\author{Jie {Zhao}\altaffilmark{1}, Brigitte {Schmieder}\altaffilmark{2}, Hui {Li}\altaffilmark{1},
Etienne {Pariat}\altaffilmark{2}, Xiaoshuai {Zhu}\altaffilmark{3}, Li {Feng}\altaffilmark{1,5}, Michalina {Grubecka}\altaffilmark{4}
}
\email{nj.lihui@pmo.ac.cn}
\altaffiltext{1}{Key Laboratory of Dark Matter and Space Astronomy, Purple Mountain Observatory, CAS, Nanjing 210008, China}
\altaffiltext{2}{LESIA, Observatoire de Paris, Section de Meudon, F-92195,Meudon Principal Cedex, France}
\altaffiltext{3}{Key Laboratory of Solar Activity, National Astronomical Observatories, Chinese Academy of Sciences, Beijing 100012, China}
\altaffiltext{4}{Astronomical Institute, University of Wroc{\l}aw, Kopernika 11, 51-622, Wroc{\l}aw, Poland}
\altaffiltext{5}{State Key Laboratory of Space Weather, National Space Science Center, Chinese Academy of Sciences, Beijing 100190, China}

\begin{abstract}
{\bf By using a new method of forced-field extrapolation}, we study the emerging flux region {\bf AR 11850 observed by the Interface Region Imaging Spectrograph (IRIS) and
Solar Dynamical Observatory (SDO). }
Our results suggest that the bright points  (BPs) in this emerging region
have responses in lines formed  from the upper photosphere to the transition region, with a relatively similar morphology. {\bf They have an oscillation of} several minutes according to the {\bf Atmospheric Imaging Assembly (AIA)} data at 1600 and 1700 \AA\,.  {\bf The ratio between the BP intensities measured in 1600  \AA\, and 1700 \AA\,  filtergrams reveals that  these BPs are heated differently.}
Our analysis of the {\it \hmi} (HMI) vector magnetic field and the corresponding topology in AR11850 indicates that the BPs  are located at the polarity inversion line (PIL) and most of them related with  magnetic reconnection or cancelation.
{\bf The heating of the BPs might be different due to different magnetic topology.}
{\bf We find that the heating due to} the magnetic cancelation would be stronger
than the {\bf case of} bald patch reconnection. The plasma density rather than the magnetic field strength could play a dominant role in this process.
{\bf Based on physical conditions in
the lower atmosphere,} our forced-field extrapolation shows consistent results  between the {\bf bright arcades visible in slit-jaw image (SJI) 1400 \AA\ and}
the extrapolated field lines that {\bf pass} through the bald patches. It
provides a reliable observational evidence for testing the mechanism of
magnetic reconnection for the BPs and arcades in emerging
flux region, as {\bf proposed in simulation works.}

\end{abstract}


\section{Introduction}

{\bf Prevalent} energy releases {\bf can occur} in the solar atmosphere in  a large range of  scales from solar flares to small energy release events.
Solar flares have been extensively studied \citep[][and the references therein]
{1981GApFD..18..332P,1991VA.....34..353P,2009AdSpR..43..739S,
2015SoPh..290.3399M,2015SoPh..290.3425J}
as it is the most energetic phenomenon and frequently  accompanied {\bf by} coronal
mass ejection, which  may influence the space weather and even our earth.
The small-scale energy release events {\bf happen} more
frequently and have been named as micro- and nanoflares,
bright points (BPs), blinkers. {\bf They have been observed by different instruments \citep{1995SoPh..156..245S,
2002SoPh..206..249P,2004psci.book.....A} in different
wavelengths from visible to soft X-ray.} They have important impacts on
{\bf the} heating mechanisms of the  solar atmosphere.
Such events  take place in different {\bf atmospheric} layers from the photosphere to the corona
\citep{1997SoPh..175..467H,2007ApJ...661.1272B,2013SoPh..286..125C},
and ubiquitous in active regions, quiet {\bf Sun} and coronal holes
\citep{1995PASJ...47..251S,2001SoPh..198..347Z,2011A&A...529A..21K}.
With  the  {\bf ground-based and space telescopes},
photospheric BPs are {\bf often studied to investigate}
the condition of the convective
flows in and below the photosphere \citep{2013Ap&SS.348...17F,2014A&A...563A.101J,
2015ApJ...810...88Y,2016SoPh..291.1089Y,2016SoPh..291..357J,2016RAA....16e...9J}.
These flows are widely accepted to be the ultimate source of the energy;
coronal BPs as well as
coronal loops are also frequently discussed
\citep{2013NewA...23...19L,2014A&A...568A..30Z,
2015ApJ...807..175A,2015ApJ...814..124C,2016ApJ...818....9M}
as they were thought to be the most potential
candidates for coronal heating.

In fact, the energy coming from the solar convection region
heats the solar atmosphere through the stressing of the magnetic field lines,
and {\bf manifests} itself as waves and current sheets leading to reconnection
\citep[see review in][]{2014SSRv..186..227S}.
The anomalous high temperature {\bf in} the corona is one manifestation
of this process. The temperature of the chromosphere is not as high
as {\bf that of} the corona because of its  higher
density. Hence, we need to study the atmosphere as an entirety  to understand
the heating process \citep[such as][]{1997LNP...489..139S,
2004ApJ...601..530S,2007PASJ...59S.643L,2012RAA....12.1681Z},
and  particularly  pay attention to the
phenomena {\bf happening} in the chromosphere and transition region.

Previous diagnostic {\bf work} have been done on
the small-scale energy release events in this interface region
\citep{1997LNP...489..139S,2004ApJ...614.1099P,2006AdSpR..38..902P,2007A&A...473..279P,
2010A&A...519A..58T,2010ApJ...724..640C,
2010ApJ...720.1472B,2013ApJ...777..135B,2015AdSpR..55..886L},
with limited wavelengths, resolution, sensitivity or {\bf observation time.}
The recently launched {\it Interface Region Imaging Spectrograph} \citep[IRIS,][]{2014SoPh..289.2733D},
which provided images and spectrum with
considerably high temporal and spatial resolution, has opened a new
window for investigating the small-scale energy release events
in the interface region. Some impressive findings
\citep[][]{2014Sci...346E.315H,2014Sci...346D.315D,
2014Sci...346A.315T,2014Sci...346C.315P,
2015ApJ...799L...3R,2015ApJ...812...11V} and
more specific works
\citep{2014ApJ...797...88H,2015ApJ...803...44M,2015ApJ...810...38K,2016A&A...586A..25P,
2016ApJ...817..124S,2016ApJ...826L..18B} have been introduced.
Most of them have focused on the transition region loops, torsional motions,
solar spicules, bright grains, explosive events, Ellerman bombs (EBs) and the hot bombs which turn out to be
the {\bf analogies} of EBs \citep{2016ApJ...824...96T}.
Such phenomena have been  studied in details, including their spectrum,
temperature, morphology, lifetime.
However, no direct analysis of the related magnetic field has been done yet.

Although it is still {\bf not clear whether waves or
reconnection} play a dominant role in heating the solar atmosphere,
it is well accepted that the solar plasma, which is ionized due to the
high temperature, is highly coupled {\bf to} the magnetic field.
{\bf An evidence is the dominance of the structured magnetic field associated with }the inhomogeneities
of the emission in the solar corona.
Hence, the magnetic topology, especially the special features
such as null points, separatrix surfaces, separators as well as quasi-separatrix layers (QSLs), is considered to be important for studying
the {\bf atmospheric} heating.
The magnetic {\bf field} associated with the BPs {\bf can} be achieved
thanks to the availability of {\bf high quality vector magnetograms}, and
the advanced reconstruction {\bf method of the coronal force-free
magnetic field.} As a result,
the coronal BPs and the associated
magnetic features have been well studied in the tenuous and highly
ionized coronal plasma where magnetic reconnection {\bf can occur easily} \citep[such as][]{2012ApJ...746...19Z,2014ApJ...794...79N}.
A bunch of {\bf work} have made some  approximations in studying the
interface brightenings with the photospheric magnetic fields
\citep[such as][]{2007A&A...473..279P,2012Ap&SS.341..215L,2015PASJ...67...40J,2016A&A...589A.114L}

Several simulation work have studied the heating of the
solar atmosphere from convection zone to the corona
\citep{2007SoPh..245...55P,2009ApJ...701.1911P,
2010ApJ...710.1387J,2012ApJ...751..152J,2015ApJ...799...79N,2015ApJ...812...92N,2016arXiv161101746N},
yet the investigation of the magnetic features {\bf in} real physical
conditions is {\bf considerably} insufficient in the lower atmosphere.
\cite{2003AdSpR..32.1875S} discussed the linear force-free-field (LFFF) method
in emerging active region and also its limitation.
In \cite{2007ApJ...671L.205F}, they even pointed out the inadequacy of the
LFFF extrapolation in the corona by comparing with their reliable
coronal loop reconstruction.
\citet{2004ApJ...614.1099P,2006AdSpR..38..902P}
studied the emergence of a magnetic flux tube and the heating {\bf during} this process.
Their LFFF extrapolation result suggests
that the impulsive heating may happen not only at the
locations of the bald patches but also at the footpoints  and along
the field lines that passing through these bald patches.
\citet{2012SoPh..278...73V} focused on the {\bf entire} process of the flux emergence,
from the first appearance of the magnetic signature {\bf in} the photosphere, until
the formation of the main bipole of the active region. They adopted the
non-linear force-free-field (NLFFF) extrapolation based on the magneto-frictional method
to construct the three-dimensional magnetic field in the emerging flux region.
\citet{2016ApJ...820L..17H} studied a micro-flare in the
chromosphere with the spectra {\bf measured by} the New Solar Telescope {\bf \citep[NST;][]{2003JKAS...36S.125G}} at Big Bear Solar Observatory (BBSO) and
extrapolated magnetic fields from {\it \hmi} {\bf \citep[HMI;][]{2012SoPh..275..207S}} {\bf with} a NLFFF approach
\citep{2004SoPh..222..247W,2010A&A...516A.107W}.
However, {\bf the force-free assumption inside these work
is probably not} consistent with the physical conditions in the lower atmosphere.
In fact, {\bf all the NLFFF extrapolation methods encounter more or less} the following difficulties in the investigation of
the magnetic field in the interface region: Firstly, the plasma in the lower atmosphere
is partially ionized and with high density. In the chromosphere the plasma beta fluctuates
a lot and can be either above or bellow 1, which
makes it much more sophisticated in analyzing the initiation and
driving mechanism of the reconnection.
Secondly, the magnetic field in the lower atmosphere, which is filled with
short and low-lying loops, is much more complex than in the corona.
Thirdly, the Lorentz {\bf force} is nonzero,
{\bf hence the force-free assumption used in the extrapolation
is invalid.}
Therefore {\bf in this paper we applied a new forced-field extrapolation method based on
a magnetohydrodynamic (MHD) relaxation method}
developed recently by \cite{2013ApJ...768..119Z}, which has already shown their
advantages in \citet{2016ApJ...826...51X}.

In this work, we investigate the magnetic topology of the BPs happening
in between an emerging flux region  observed {\bf by}  the {\it Solar Dynamical Observatory} (SDO) and IRIS.
{\bf A new method} of the forced-field extrapolation
is adopted \citep{2013ApJ...768..119Z}. This paper is organized as follows:
the observations are introduced in Section 2 and extrapolation in Section 3.
We discuss our results and  make conclusions in Section 4.

\section{Observations}
\subsection{Overview of the Active Region}
NOAA AR11850 appears at the location of E$24^{\circ}$N$8^{\circ}$ on September 24, 2013. It {\bf is an emerging flux region}
and {\bf consists of three dispersed spots} (Figure \ref{fig00}a).
The AR11850 has been captured by telescopes both in the space and on the ground, which makes it a
suitable region {\bf for analyzing} the successive heating of the solar atmosphere.
There are many BPs appearing in between this active region,
\citet{2014Sci...346C.315P} studied four of them and considered them as "hot explosions";
\citet{2016A&A...593A..32G} made statistics {\bf on} the compact BPs observed in this region.

The {\it \aia} (AIA) instrument {\bf \citep{2012SoPh..275...17L}} onboard SDO satellite obtains full-disk images in UV and EUV wavelengths {\bf and monitors} the
atmosphere from the chromosphere to the corona with high spatial resolution (0.6 arcsec per pixel) and continuous temporal coverage.
The {\bf IRIS mission {\bf \citep{2014SoPh..289.2733D}} obtains UV spectra and images of} the
chromosphere and transition region, with a spectral resolution of respectively  $\approx26$ and $\approx52$ m\AA\
according to the wavelength range,
and a spatial resolution of 0.33 -- 0.4 arcsec for the images.
{\bf IRIS observed AR11850 with a FOV of 141 arcsec $\times$ 175 arcsec for the raster and 166 arcsec $\times$ 175 arcsec for the image.}

The cadence of the slit-jaw  images (SJI)  in the {\bf Si IV} 1400 \AA\ filter  is 12 {\bf seconds}.  Images using the rasters can be obtained in 20 minutes (first interval is 11:44 -- 12:04 UT and second interval 15:39 --15:59 UT) on September 24, 2013. We will mainly use the
SJI  1400 \AA\ observed at 11:52 UT and the spectrogram  in the Mg II h line wing at 2803.5 +1  \AA\  {\bf scanned from 11:50 to 12:01 UT} in  the first time interval  of observations and covering  only the emerging flux region.


\subsection{Observations {\bf by} IRIS and SDO/AIA}
Since we are mostly interested in the initial phase of the emerging process when the flux tubes
pass through the lower solar atmosphere, especially through the photosphere, we mainly
investigate {\bf the observations of this emerging flux region in UV wavelengths.}

\begin{figure}
\centering
\includegraphics[scale=1.0,trim = 2.0cm 3.0cm 0.0cm 0.0cm]{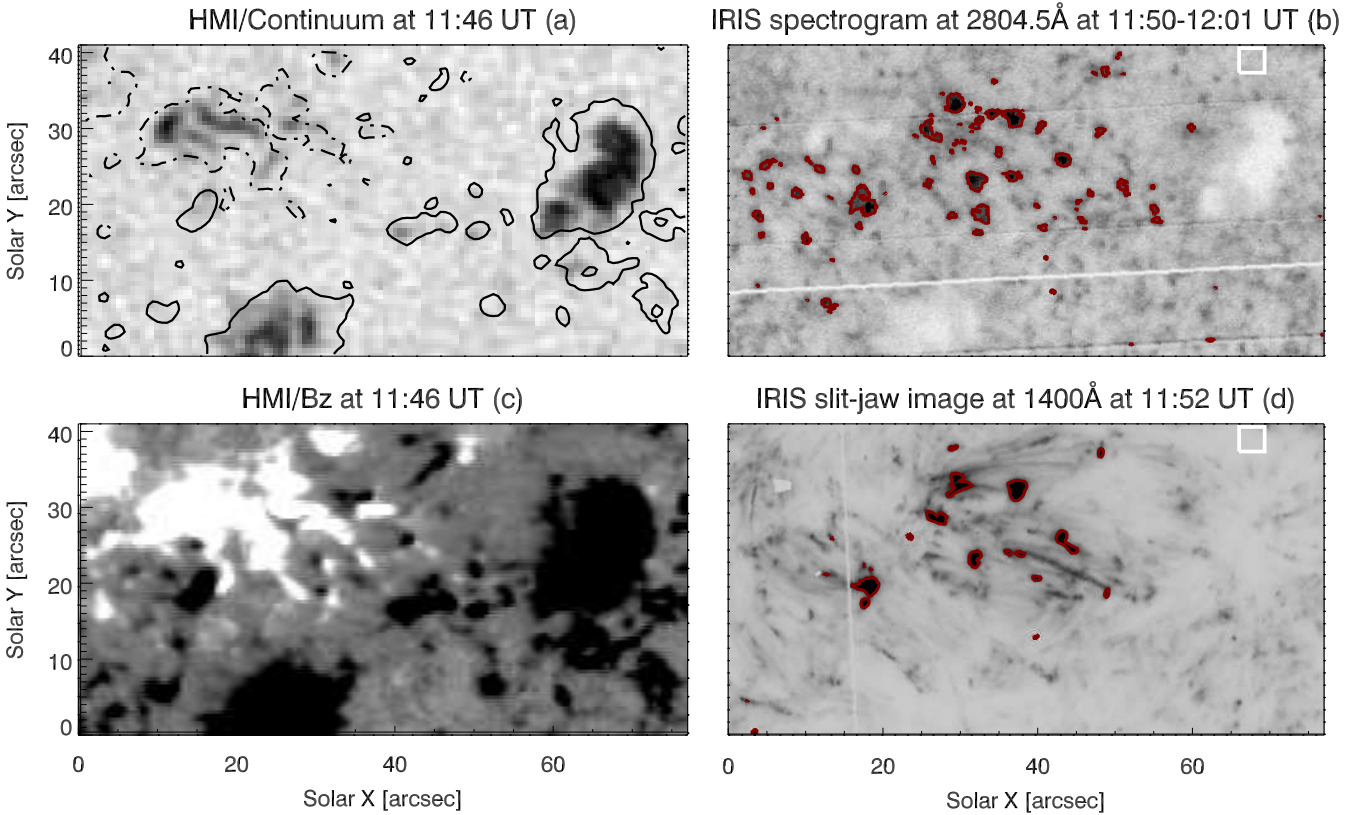}
\caption{Emerging flux with BPs
observed by IRIS and HMI in AR11850 on September 24, 2013. (a) HMI continuum image, showing the location of the spots.
Contours (values equal $\pm 200$ G, solid lines for negative polarity and dash-doted lines for positive polarity) of the vertical magnetic field are overlaid.
(b) Spectrogram between 11:50 and 12:01 UT in  Mg II h line wing  (2803.5  +1 \AA) with  red contours overlain;
(c) Vertical magnetic field component map at 11:46 UT in the  photosphere, the maximum and minimum value of the magnetic field is $\pm 200$ G.
(d) {\bf SJI} at Si IV 1400 \AA\ with red contours overlain;
We present negative color for panel (b) and (d) that the black color shows high intensity and white color for low intensity.
Same formula is used for Figure \ref{fig01} (a) and (c), Figure \ref{fig02} (a), Figure \ref{fig11} and Figure \ref{fig12}.}
\label{fig00}
\end{figure}

{\bf For a context of AR 11850, we} present the HMI continuum intensity and photospheric magnetogram  in Figure \ref{fig00} (a) and (c) to show the magnetic configuration in this region.
The images of different UV wavelengths are shown in Figure \ref{fig00} (b) and (d),
in IRIS spectrogram at 2804.5 \AA\ {\bf from 11:50 to 12:01 UT} and SJI 1400 \AA\ at 11:52 UT,
with  approximated formation temperatures of $10^{4.0}$ and $10^{4.8}$ kelvin, respectively.
In both UV images, there are several well-pronounced BPs in the moss between the magnetic polarities, which are outlined by red contours.
A bunch of plasma loops can also be recognized in 1400 \AA\ image, which might correspond to the arch filament system (AFS)
in H$\alpha$ \citep[refer to ][]{2016A&A...593A..32G}, with their feet rooted at some of the {\bf aforementioned} BPs.
In SJI 1400 \AA\, there are 19 bright points {\bf between magnetic polarities in} this active region
with a threshold value of 20 times above the $I_{smean}$, while in 2804.5 \AA\,
there are 50 bright points with a threshold value of 1.6 times above the $I_{mmean}$.
The values of $I_{smean}$ and $I_{mmean}$ represent the mean intensity {\bf in} a relatively quiet region
{\bf indicated by }the rectangular white box {\bf in} Si IV and Mg II images, respectively.

Most of the  intense BPs that appear in 2804.5 \AA\ have  a counterpart in SJI 1400 \AA\,
although they are not outlined by red contours due to the contour level that we have selected to {\bf avoid mixing} of BPs and bright loops.
In the magnetogram, we have {\bf increased} the contrast to {\bf enhance the small-scaled dipoles between the spots.}
The relationship between the bright points and magnetic dipoles will be studied in detail later.

\begin{figure}
\centering
\includegraphics[scale=1.0,trim = 0.4cm 0.0cm 0.0cm 0.0cm]{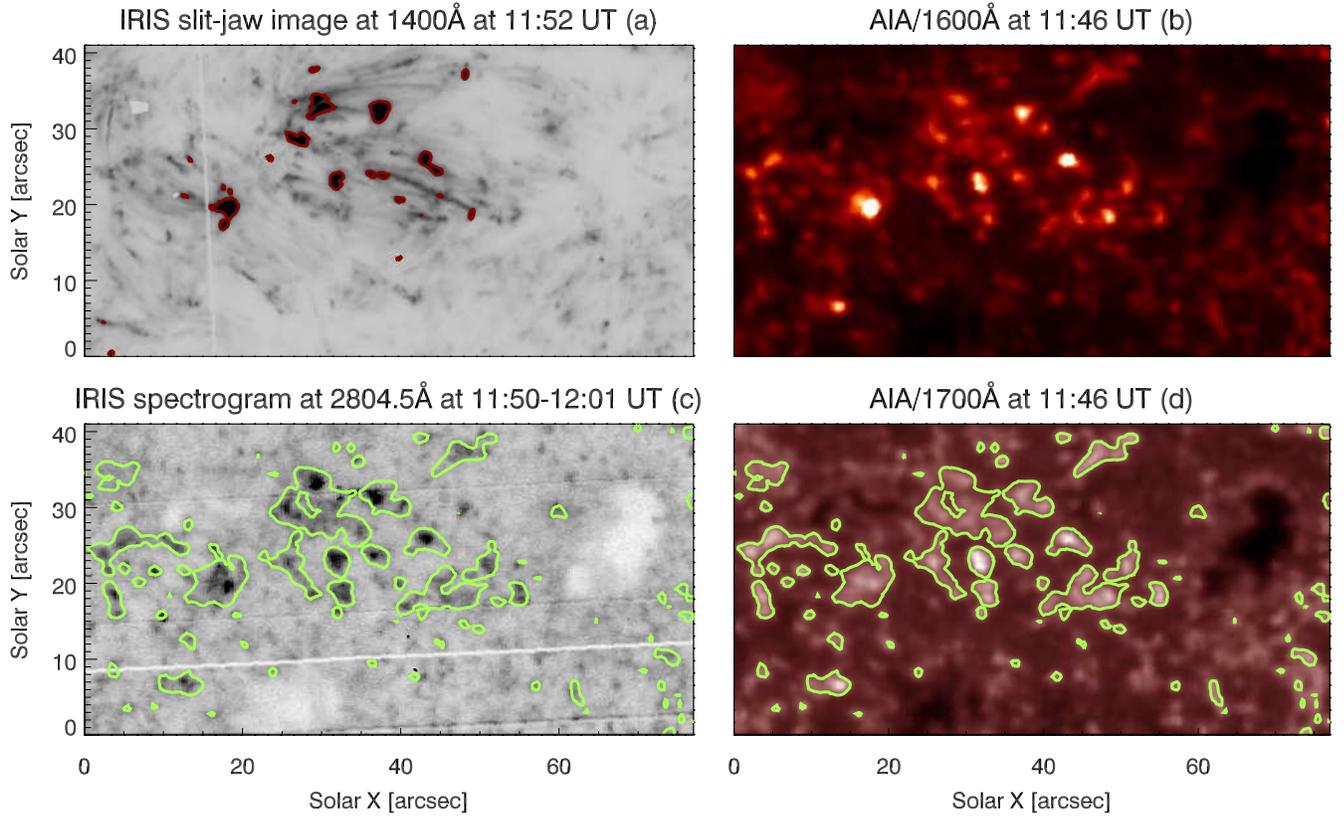}

\caption{Emerging flux with BPs
observed by IRIS and AIA in AR11850. (a) {\bf SJI} at Si IV 1400 \AA\ with contours of  the bright points;
(b) Intensity image at 1600 \AA\, (c) Spectrogram between 11:50 and 12:01 UT in Mg II  h  wing  (2803.5  +1 \AA)  with contour of 1600 \AA\ intensity image overlain;
(d) Intensity image at 1700 \AA\ with contour of 1600 \AA\ intensity image overlain.}
\label{fig01}
\end{figure}

{\bf The UV images at other wavelengths, such as AIA 1600 and 1700 \AA\, are shown in Figure \ref{fig01}.
The 1700 \AA\ emission (UV continuum) forms around the temperature minimum region of $10^{3.7}$ kelvin, while the 1600 \AA\ emission contains
UV emission like 1700 \AA\ and also the emission of C IV line 1548 \AA\ formed in the transition region temperature.}
{\bf Complementary images from IRIS are also displaced here.}
{\bf These four images reveal the atmosphere response to the magnetic flux emergence from photosphere to the transition region}.
The images from the different satellites are coaligned through the  alignment of specific features.
The contours of 1600 \AA\ intensity image are overlaid on IRIS 2804.5 \AA\ and AIA 1700 \AA\ images. {\bf It implies} that all the images are {\bf coaligned} quite well.
A large number of BPs can be identified in the moss between the spots in
1600 and 1700 \AA\ images.

\begin{figure}
\centering
\includegraphics[scale=1.0,trim = 1.0cm 0.0cm 0.0cm 0.0cm, clip]{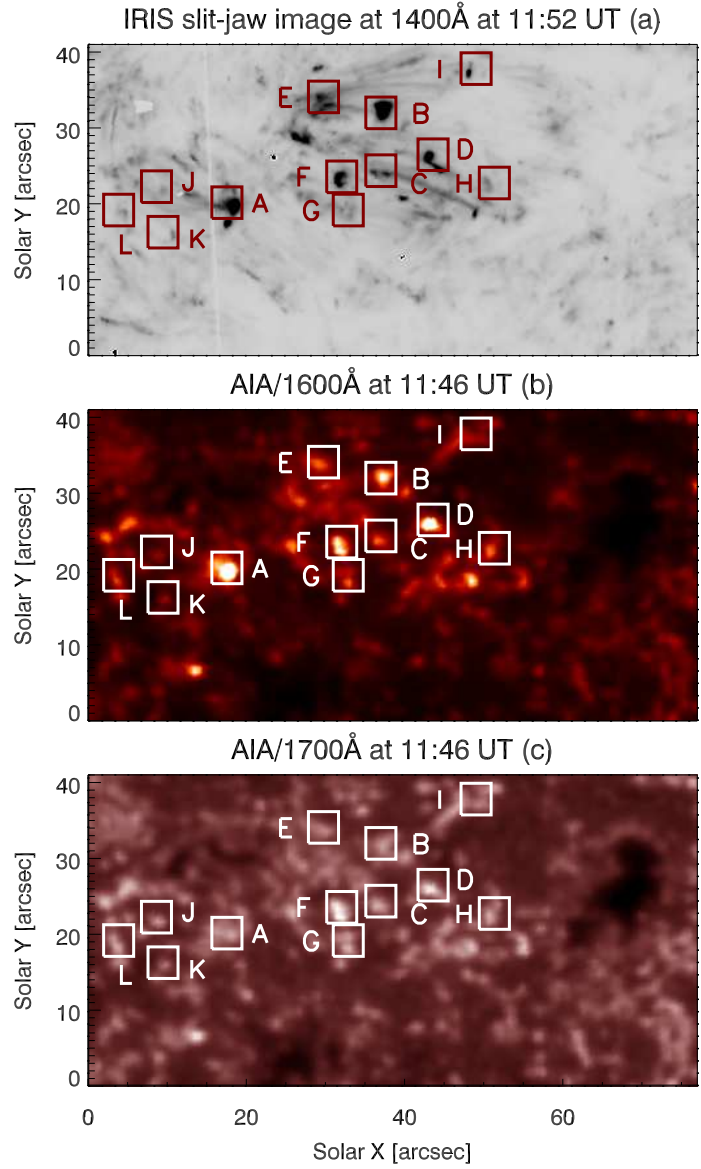}
\caption{Intensity images of AR11850 at different wavelengths at 11:52 UT and at 11:46 UT.  The square boxes select twelve  BPs (A to L)
that are  studied in detail in the paper.}
\label{fig02}
\end{figure}

\begin{figure}
\centering
\includegraphics[scale=0.8]{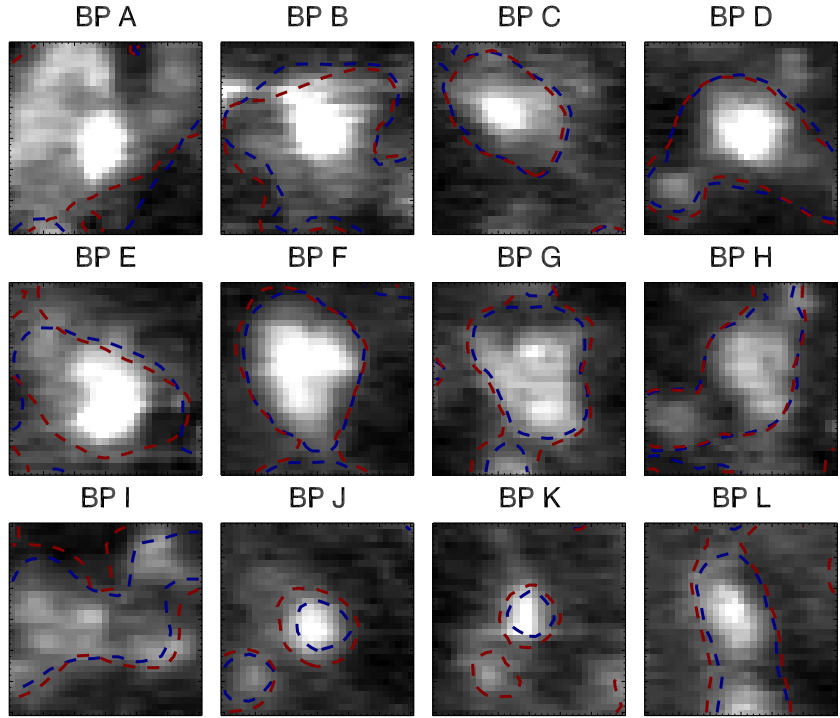}
\caption{Intensity images for BPs A -- L  in Mg II  h  wing  (2803.5  +1 \AA). The AIA 1600 \AA\ (blue) and 1700 \AA\ (red) contours are overlaid.}
\label{fig09}
\end{figure}

Among these BPs, we selected several brightest ones A-I by {\bf visual inspection of their brightness} in IRIS 1400 \AA\
and present them with rectangular box in Figure \ref{fig02}. All the selected BPs have their counterparts in AIA 1600 \AA\ and 1700 \AA. This result suggests
that the BPs are heated in the atmosphere from the photosphere to the transition region.
In order to compare the morphology of these BPs,
we overlay the contours of AIA 1600 \AA\ (2.2 times above the mean value) and  AIA 1700 \AA\ (1.4 times above the mean value)
on the spectrogram of 2804.5 \AA\ for every single BP
in Figure \ref{fig09}. We notice that most of the BPs have a similar morphology, with a more diffuse shape in the AIA observations.
{\bf As mentioned in \citet{2015ApJ...812...11V} and \citet{2016ApJ...824...96T}, the BPs observed in Mg II line wing commonly have a counterpart in the wings of H$\alpha$ like the EBs. It indicates that the BPs marked in the boxes maybe associated with EBs.}

We present the evolution of selected BPs in 1600 \AA\ in Figure \ref{fig03} and in 1700 \AA\ in Figure \ref{fig04}.
The images record the BPs evolution for 16 minutes from 11:46 UT to 12:04 UT. The black contours outline the BPs
in this rectangular region and the black arrow points out the specific one that we are interested in.
The morphology of BP A and B seems to be different in 1600 \AA\ and 1700 \AA\, while the others are similar.
The shape of these BPs always changes during the 18 minutes.
The white rectangular boxes show the times when the slit of IRIS scans through these BPs.
We calculated the intensity ratio between 1600 \AA\ and 1700 \AA\ at these specific times (Table \ref{Table1}).
{\bf All ratios are greater than unity except for BP I.} These ratios imply  that there is a  contribution of C IV  emission in the BPs.
The more the ratio is greater than unit, the more the contribution comes from C IV. Hence, we could find that the C IV  emission contributes much more
in BPs  A , B  and E, less for  C, D, F  and almost no for the others (G to L). {\bf It} means that the temperature of the BPs  A to F is higher than that of the others.
This is consistent with the results of \cite{2016A&A...593A..32G}, {\bf who} found signatures in Si IV spectra only for the BPs corresponding to the classes of A, B, D.
From the temporal evolution of the above BPs, we obtain the intensity curves in 1600 and 1700 \AA\ {\bf shown in Figure \ref{fig05}.}
{\bf The BPs appear to have an oscillation of several minutes, just like the jets.
This is probably due to the recurrent reconnection modulated by the plasma motion and releases energy quasi-periodically.
This scenario has been supported by the theoretical \citep{2007ASTRA...3...29S,2010ApJ...714.1762P} and observational \citep{2003A&A...398..775M,2004A&A...418..313U,2006A&A...446..327D,2011ApJ...732L...7Y,
2012ApJ...746...19Z,2013A&A...555A..19G,2015ApJ...806..172S,2016AN....337.1024I} work.}

\subsection{Observations {\bf by} SDO/HMI}
As the magnetic field has intimate relation with the phenomenon happened in the solar atmosphere,
our study also included the analysis of the  vector magnetic field at the photosphere as well as the extrapolated
field based on it. The vector magnetic field is provided by the HMI onboard SDO, with a spatial resolution of
0.5 arcsec per pixel and a temporal resolution of 12 {\bf minutes}.
We present the images of the  vertical component   B$_z$ and the vector  B$_{vec}$  of the observed magnetic field
in Figure \ref{fig06} (a, b, c), {\bf and display the bottom layer} of the extrapolated magnetic field in  Figure \ref{fig06} (e).
We see {\bf magnetic fields with mixed polarities} between the spots. These are the preferential places for {\bf a specific bald patch topology structure} (red points in Figure \ref{fig06} (c -- f)).
{\bf This conception dates back to \cite{1993A&A...276..564T}. The authors emphasized}
a place on the polarity inversion line (PIL) at the photosphere where
magnetic field line threading through it horizontally
from negative to positive polarity, and named it as 'Bald Patch'.
It relates to a serpentine field line when the flux tube emerging
from the convective region to the solar atmosphere
\citep{2001ApJ...546..509F,2001ApJ...554L.111F,2009ApJ...701.1911P}.
Similar structure has also been found in the low altitude
atmosphere and can be identified as 'magnetic dip' refer to \cite{2004ApJ...614.1099P}.
Cool and dense material {\bf often deposite in these
dips.}
Such location as well as its related BP separatrix has been found to be preferential
place for reconnection {\bf when proper surface flows are involved. Such reconnection} is responsible for the EBs and brightening loops in many studies of observation and simulation
\citep{2009ApJ...701.1911P}.
In the spectral observations, {\bf blueshift and redshift of spectral lines (such as Si IV, C II, Mg II) associated with the bidirectional flows are frequently recognized around these places \citep{2015ApJ...804...82C}.}
The comparison of the images of 2804.5 \AA\, overlaid by  bald patches  derived from observation and from extrapolation,   shows a similar distribution of the bald patches in the field of view (Figure \ref{fig06} (d, f)). It means that the extrapolation correctly retrieves the magnetic field structures.

The calculation of bald pathes or magnetic dips in 3D volume is  based on the magnetic field in 3D. {\bf It is provided
by the forced-field extrapolation described in the next section.}

\section{Forced-Field Extrapolation}

To investigate the relationship of the BPs and loops with
the associated magnetic field, we extrapolated the magnetic field from
Space-weather HMI Active Region Patch (SHARP) Cylindrical Equal Area (CEA)
data of SDO/HMI. {\bf The forced-field extrapolation code is described in \cite{2013ApJ...768..119Z}
and computes the magnetic field by
solving full MHD equations using a kind of relaxation method.} The initial state
comprises a plane-parallel multilayered hydrostatic model \citep{2001ApJ...546..509F},
embedded with a potential magnetic field \citep{1982SoPh...76..301S} determined by the normal
component of the vector magnetogram. The outflow condition is
applied for both sides and upper boundaries. At the lower boundary the normal
component of the magnetic field is  fixed, while the transverse field
is slowly changed from the initial condition to the observed field.
This is called the "stress and relax" approach \citep{1996ApJ...473.1095R} which drives
the system to evolve. Finally, the Lorentz Force near the photosphere
can be balanced by the pressure gradient and plasma gravity, and the forced
equilibrium of the {\bf entire} region can be reached. Our extrapolation
is done on the initial size of the grid, e.g. 1 grid equal 0.5 arcsec, with
a domain of 328 $\times$ 248 $\times$ 160 grid.

Then we calculated the locations of bald patches referring
to the equation in \cite{2004ApJ...614.1099P}:
\begin{equation}
B_{z} = 0 \quad and \quad {\bm B}\bullet{\nabla}{B_{z}} > 0
\end{equation}
{\bf the magnetic topologies of BPs A -- L
in detail are illustrated in Figure \ref{fig08}.}
In our study, the BPs A, B and I are located above {\bf the magnetic PIL where
magnetic flux cancelation is dominant,
the BPs D, F, G, J have consistent locations with their corresponding bald patches.
However, we don't see the consistency of the locations between the BPs C, H, K, L and their corresponding bald patches.}
The BP E is located over the PIL,
without bald patch or cancelation topology. {\bf In our later study, we find that BPs C, E, H, K, L are at the footpoints of bald patch
related separatrix layers. The relevant bald patches are located far away from the BPs.}

{\bf The temporal evolution of the magnetic flux of these BPs is shown in Figure \ref{fig10}.
All these curves start from 11:09 UT and end at 12:09 UT. They cover the time range of the AIA and IRIS observations indicated by the
grey rectangular region (11:46 UT to 12:04 UT) in each panel. The curves of BPs A, B and D, F display a continuous decrease, while the curves
of BPs I and G, J show a temporary one during the observation (in the grey region).
It manifests that magnetic cancelation and reconnection, which are responsible for the plasma heating, have happened either continuously or temporarily. Apparently,
the continuous cancelation or reconnection would produce stronger heating effect than the temporary one according to Table \ref{Table1}.
The rest curves describe the local condition of the magnetic flux, and are meaningless for understanding the BPs C, E, H, K, L that are related with separatrix layer footpoints.
The different magnetic situation of the BPs is clearly shown in Figure \ref{fig07}}.

Considering that the BPs A, B, D and I correspond to
bombs 3, 4, 1 and 2 in \citet{2014Sci...346C.315P} respectively, the
bomb 1 (BP D)  seems to be related with bald patch reconnection while the
other three bombs 2  (BP I), 3 (BP A),  4  (BP B) are {\bf produced} by flux cancelation.
In addition, \cite{2016A&A...593A..32G} studied the formation height
of the same BPs and listed their result in their Table 3.
The altitude of hot spots A -- E  ranges from 75 -- 900 km
from a 1D solar atmosphere model with the radiative transfer code of
\cite{1995A&A...299..563H}. The formation height
of A,  B is relatively higher than D, which could indicate that
the cancelation reconnection happened a little
higher than the BP reconnection.
It is surprising to detect that the "hot explosions"  (BPs A, B and I ) of \citet{2014Sci...346C.315P} correspond principally {\bf to the} flux cancellation region {\bf rather than} a bald patch region like the bomb 1  (point D in our analysis).  Their large emission {\bf values indicate} that reconnection by flux cancellation lead to stronger heating than in bald patch configuration.

The extrapolated field lines lying between the polarities and passing through
the bald patches which are labeled as red points
are {\bf overplotted on the IRIS {\bf SJI}
1400 \AA\ in Figure \ref{fig11} and the upper panel of Figure \ref{fig12}.} Figure \ref{fig11}  shows an overview of the field lines in this AR, while in Figure \ref{fig12}
{\bf we only selected several representatives for clarity.}
These field lines have diverse lengths but with a coherent direction
{\bf going} from the positive to the negative polarities.
Most of the brightening loops appearing in 1400 \AA\ have good correspondence with
the magnetic field lines in profile. Hence, we suggest that the brightening
loops in the interface layer of this active region
are at least partially {\bf contributed} by the reconnection along the bald patch separatrices.

We also extract these field lines and exhibit their side views in the bottom panels {\bf of} Figure \ref{fig12}.
The spatial distribution of the sea-serpent structures is {\bf prominent}.
Most of these structures have relatively low altitude, while some higher {\bf field} lines have a height of less than 3.5 Mm.
The white points mark the locations of the bald patch that the field lines passing through.
It manifests that the bald patch often connect two arcades, one is lower and shorter and the other
is higher and longer, just like the case from MHD model in \cite{1998SoPh..183..369A}.

\section{Discussion and Conclusion}

In this work, we have investigated an active region {\bf during} its emerging phase. {\bf In} this phase, there are plenty of
BPs that appear in between this region, i.e. in the moss, which indicates that the {\bf atmosphere above} has been heated. Hence,
it {\bf is} meaningful to study the properties of the BPs
for understanding the heating of the upper atmosphere under the following two questions: 1) {\bf To} what extent does the emerging flux heat the {\bf atmosphere above}? and 2) At which location does the heating become more effective?

This active region has been observed by the {\it Multichannel Subtractive Double Pass spectrograph} (MSDP) in the Meudon Solar Tower on the ground and IRIS as well as SDO in the space. These telescopes provide the images of this region from photosphere to the corona, and also the {\bf spectra} of the transition region which has already
been analyzed by \cite{2014Sci...346C.315P} and \cite{2016A&A...593A..32G}.
Here we selected IRIS {\bf SJI} at
1400 \AA\ and spectrogram in Mg II h line wing at 2804.5 \AA\ {,} AIA 1600 \AA\ and 1700 \AA\ to
{\bf analyze} the atmosphere response to the emerging flux.
Our results {\bf demonstrate} that the BPs that appear in the IRIS {\bf SJI} 1400 \AA\ have their {\bf counterparts}
in other wavelengths {\bf (formed at the minimum temperature)} that we mentioned above, and most of them have similar morphology.
{\bf Referring} to the formation temperatures of these spectral lines, we suggest
that the emerging flux could heat the solar atmosphere from the upper photosphere to the transition region.

\begin{figure}
\centering
\includegraphics[scale=0.8]{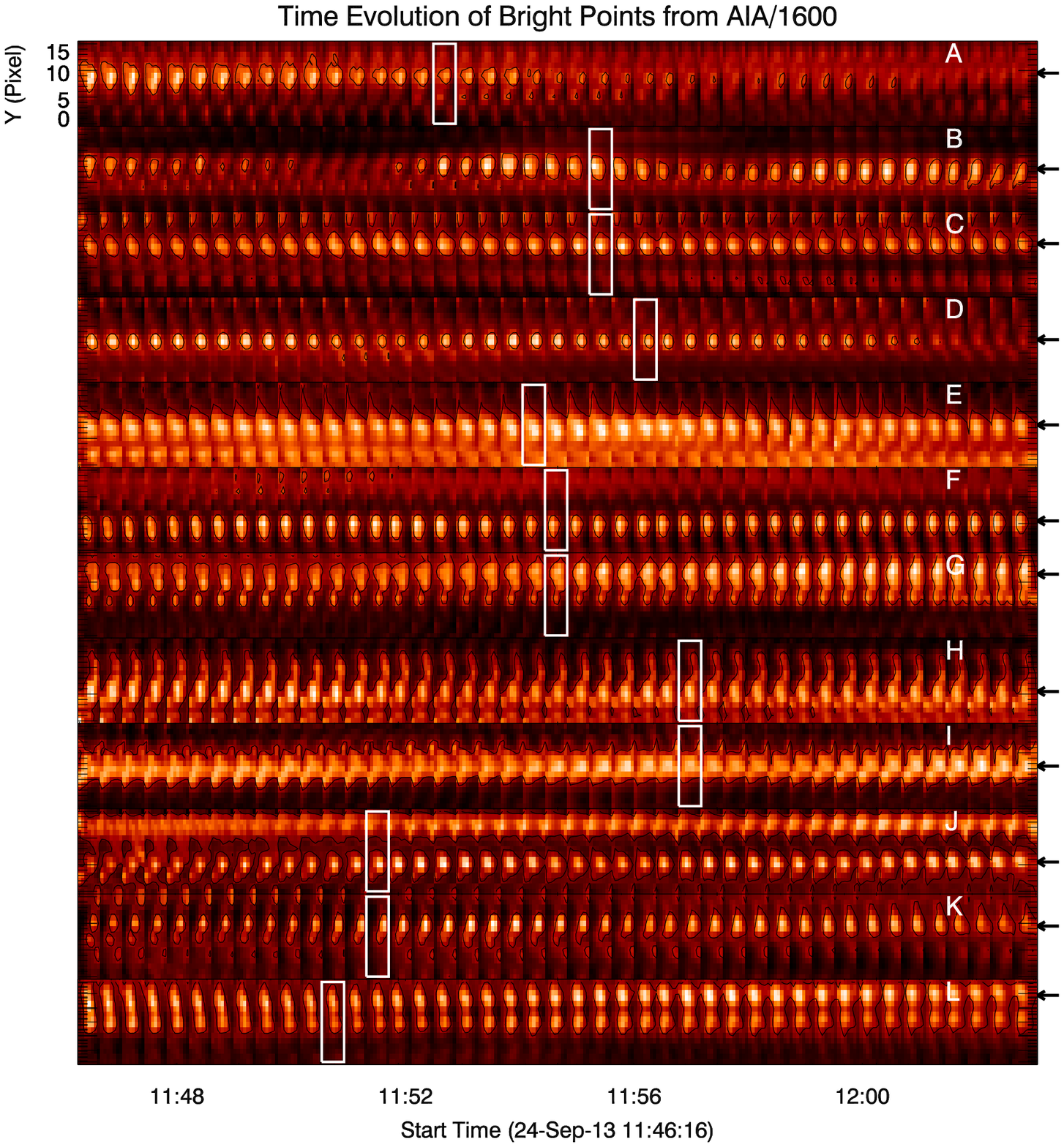}
\caption{{\bf Temporal evolution }of the brightness in 1600 \AA\ of  the twelve selected BPs. The boxes indicate the time of the BP spectra obtained by IRIS during the raster.}
\label{fig03}
\end{figure}

\begin{figure}
\centering
\includegraphics[scale=1.0]{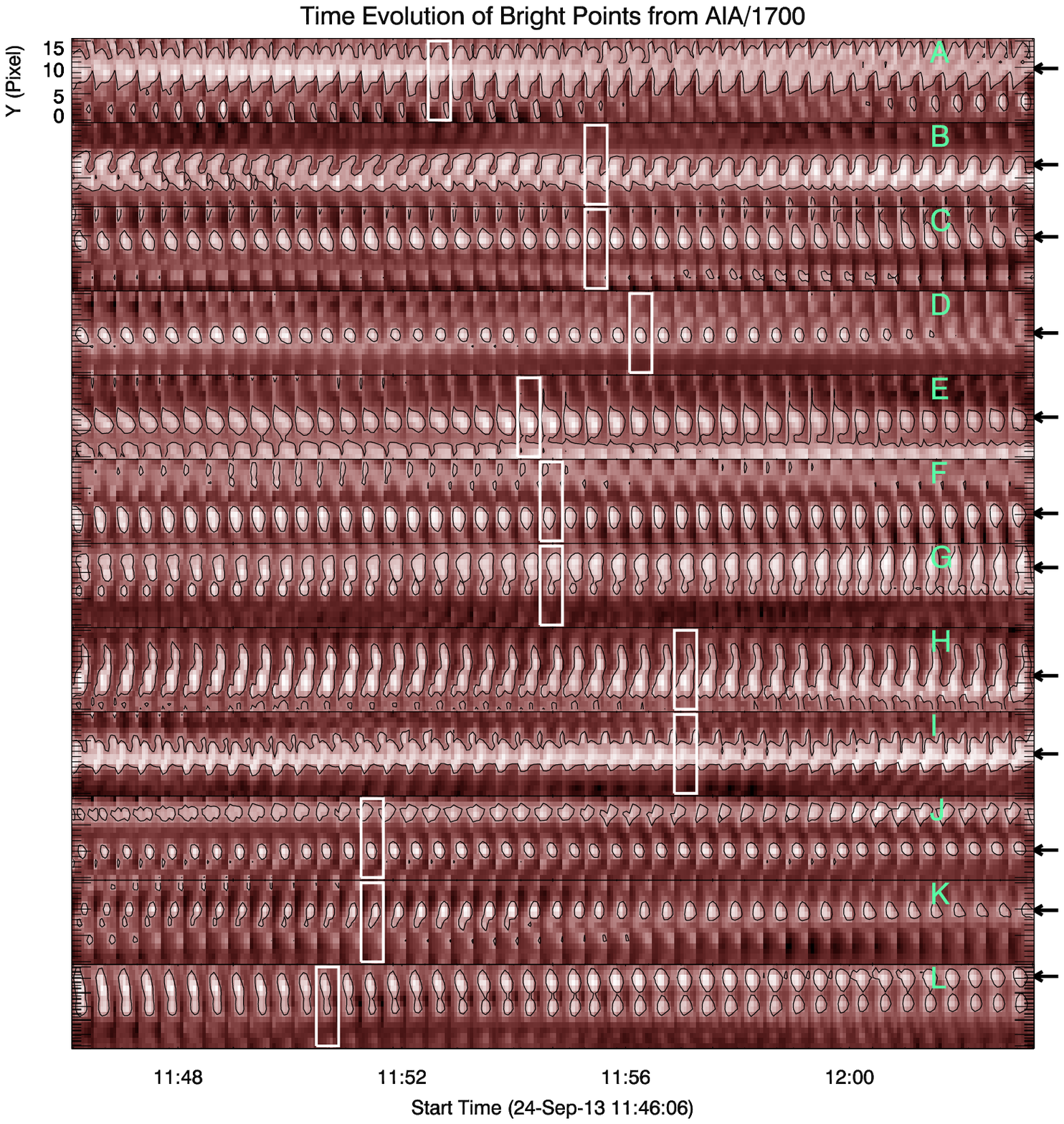}
\caption{{\bf Temporal evolution }of the brightness in 1700 \AA\ of twelve selected BPs. The boxes indicate the time of the BP spectra obtained by IRIS during the raster.}
\label{fig04}
\end{figure}

\begin{figure}
\centering
\includegraphics[scale=0.8,trim = 0.0cm 0.0cm 0.0cm 0.5cm]{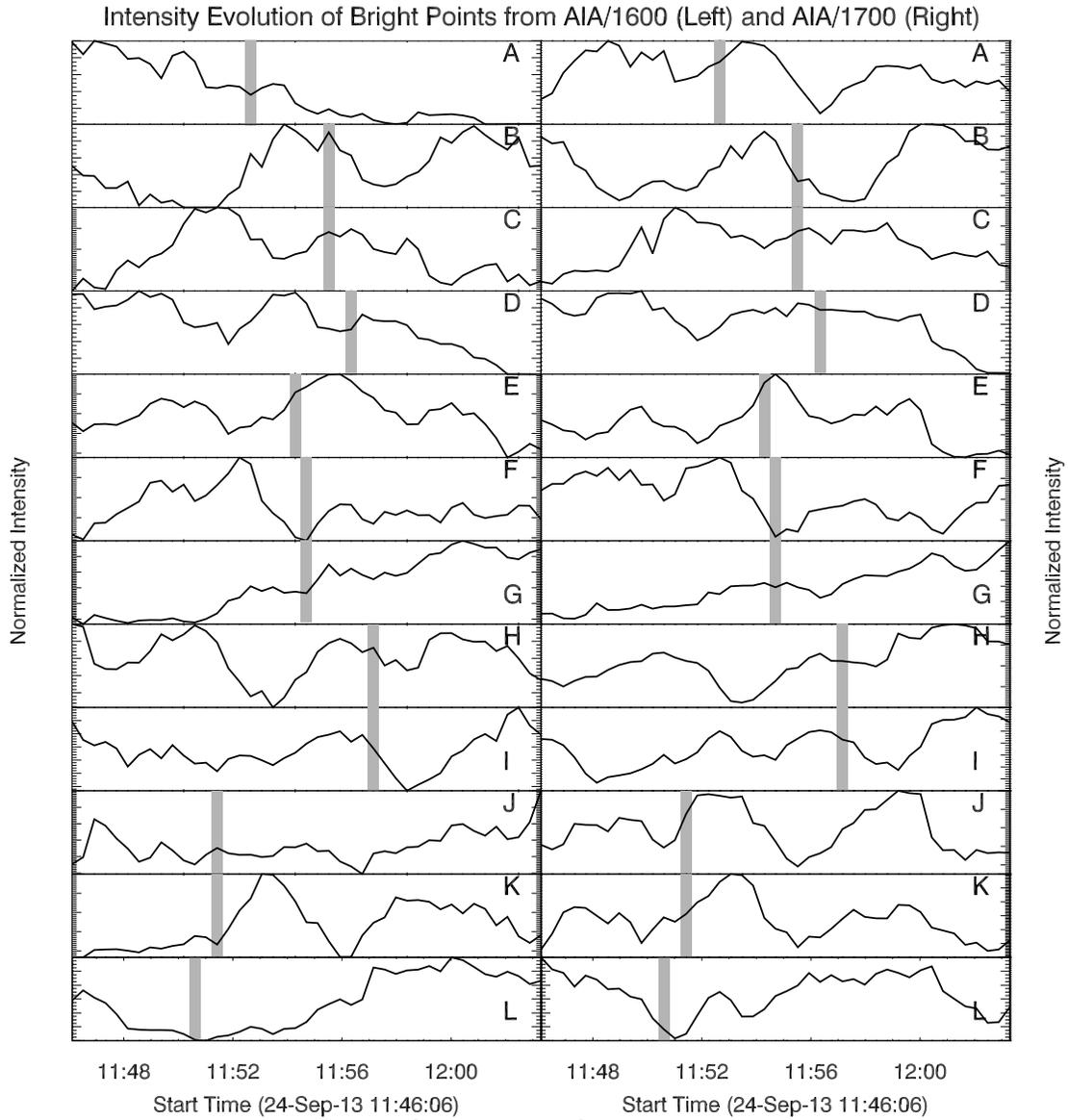}
\caption{Intensity curves in 1600 \AA\ and 1700 \AA\ of twelve selected BPs versus time. The {\bf grey} rectangular regions label the {\bf IRIS}
scanning time of the BPs, respectively}
\label{fig05}
\end{figure}

\begin{figure}
\centering
\includegraphics[scale=0.8]{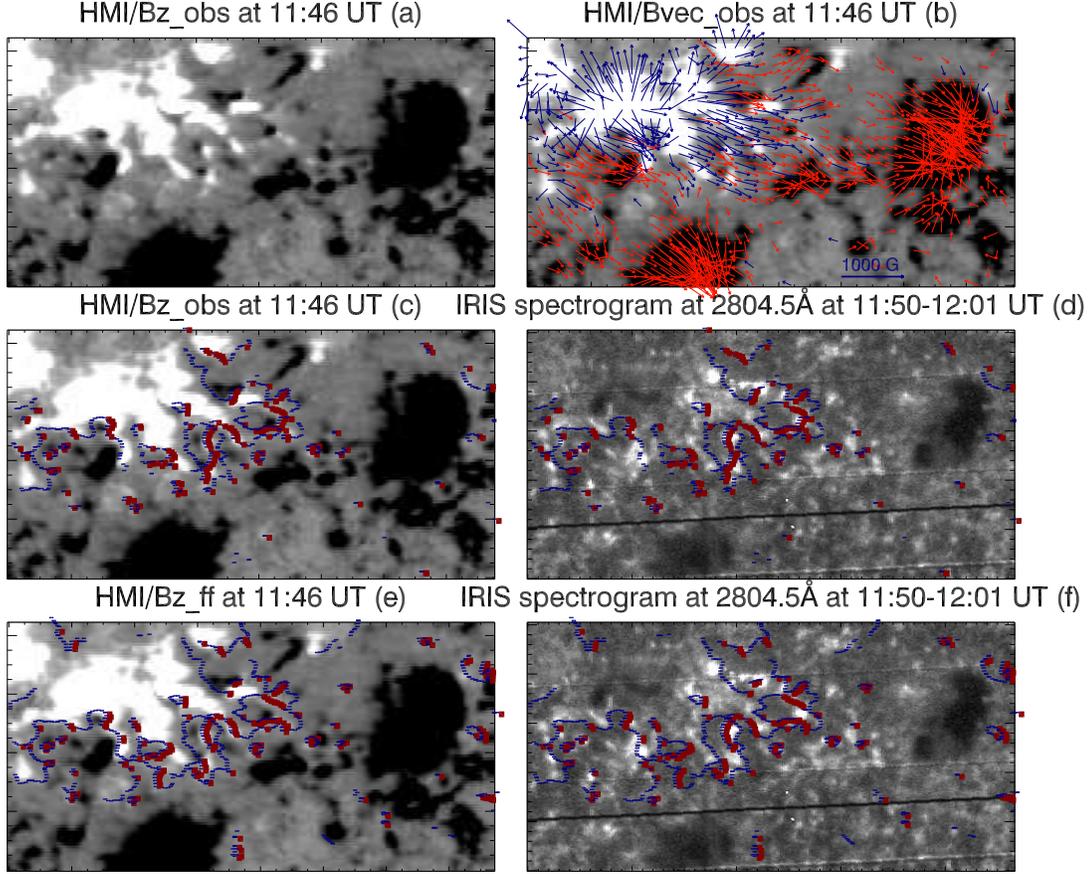}
\caption{Magnetograms of the  photosphere at 11:46 UT and IRIS spectrogram between 11:50 UT and 12:01 UT.
(a) Vertical magnetic field component  at  the photosphere from observations;
(b) Vector magnetogram at photosphere. The background is vertical magnetic field and the blue and red arrows represent the  horizontal magnetic field with positive and negative vertical magnetic field, respectively;
(c) Same as (a) but with PIL (blue dots) and bald patches (red dots) overlain;
(d) Spectrogram at Mg II 2804.5 \AA\ with bald patches from observations located;
(e) Vertical magnetogram at photosphere from forced field extrapolation with PIL (blue dots) and bald patches (red dots) {\bf overlaid};
(f) Spectrogram at Mg II 2804.5 \AA\ with bald patches from extrapolation located. }
\label{fig06}
\end{figure}

\begin{figure}
\centering
\includegraphics[scale=0.8, angle=0, trim = 0.5cm 3.0cm 1.0cm 0.0cm, clip]{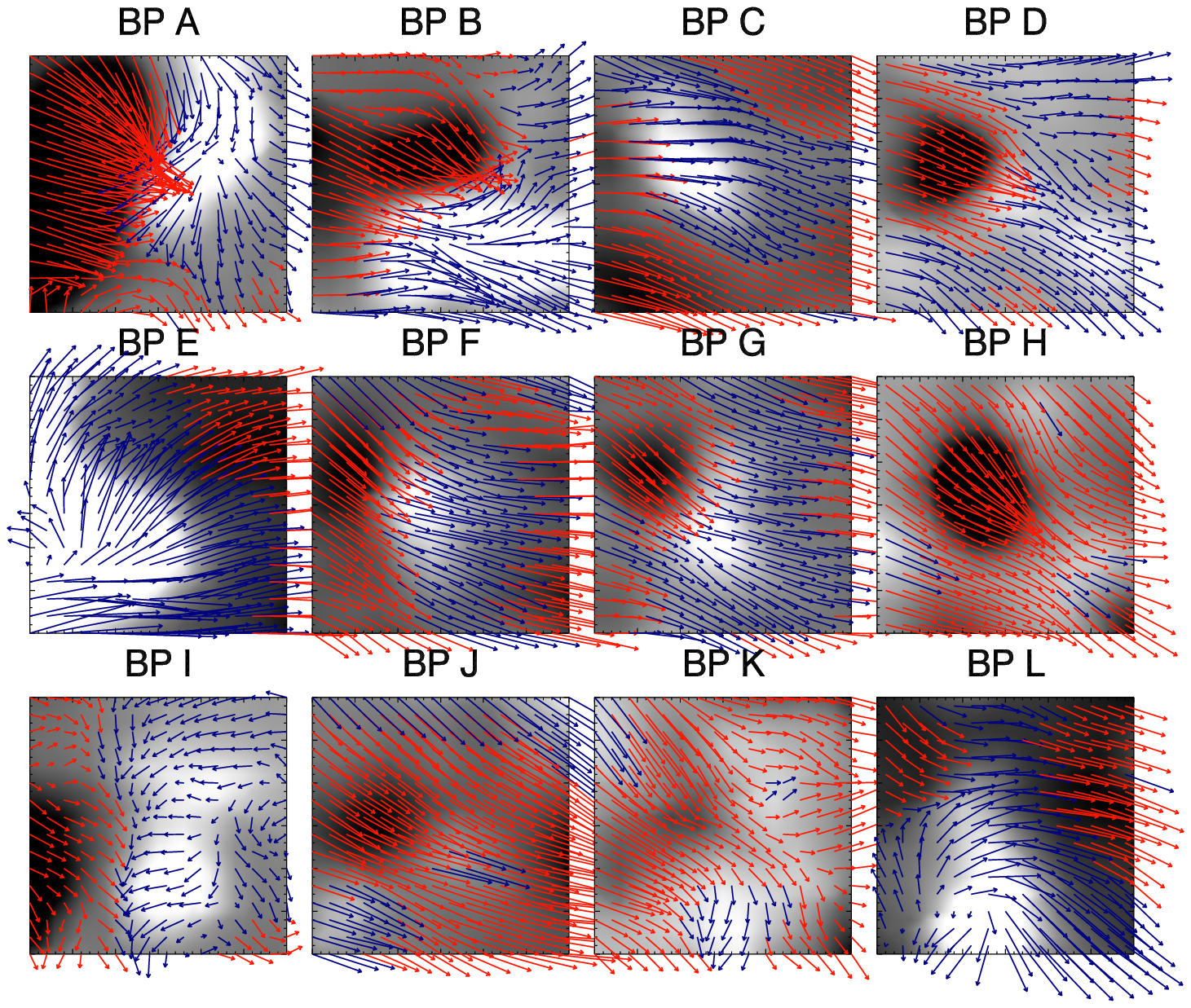}
\includegraphics[scale=0.8, angle=0, trim = 0.5cm 2.0cm 1.0cm 0.0cm, clip]{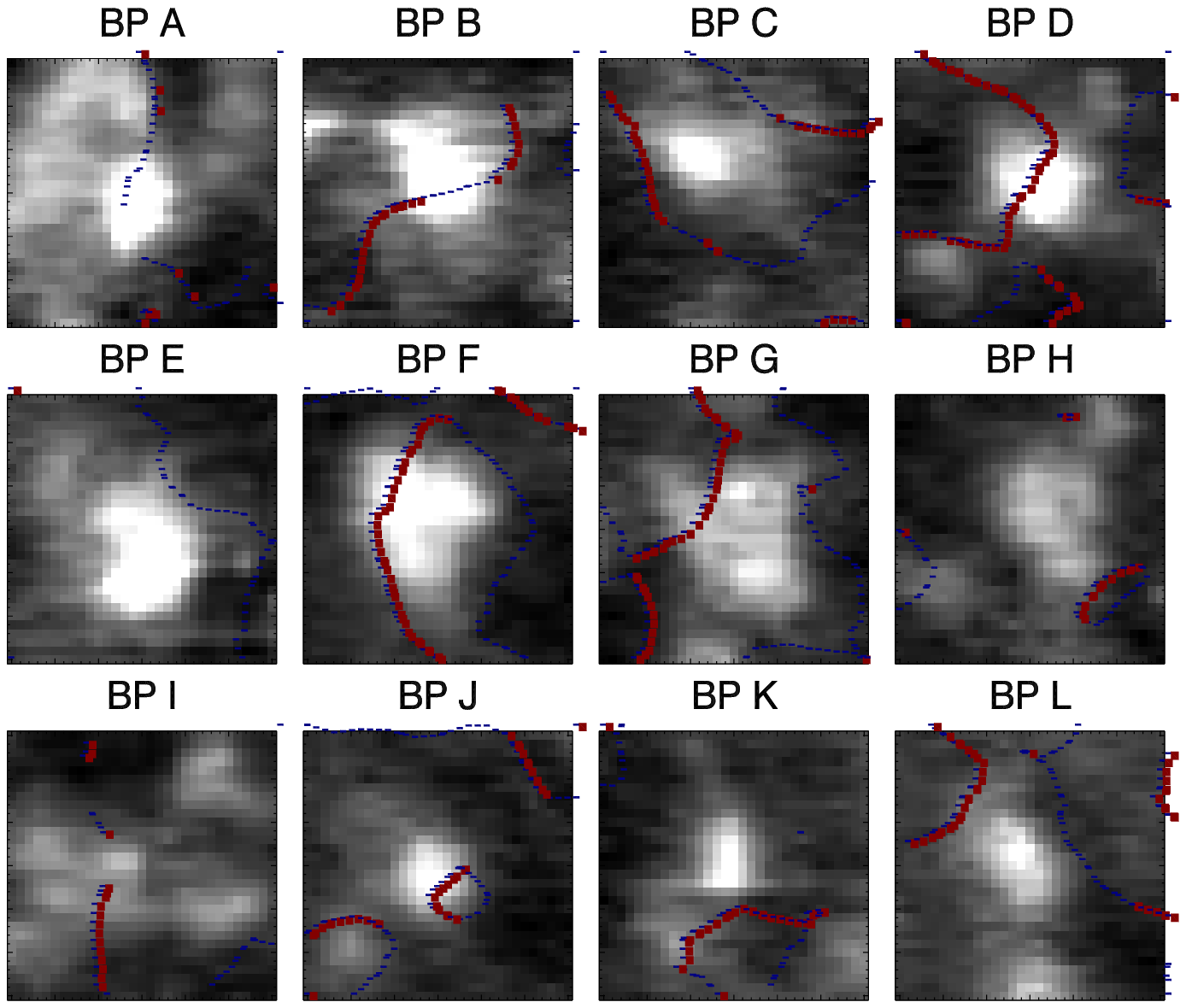}
\caption{Upper three rows show the vector magnetic field of selected BPs and lower rows show the locations of  PIL  (blue dots) and bald patches (red dots) on the IRIS Mg II  h  wing  (2803.5  +1 \AA) images.}
\label{fig08}
\end{figure}

\begin{figure}
\centering
\includegraphics[scale=0.8, angle=0]{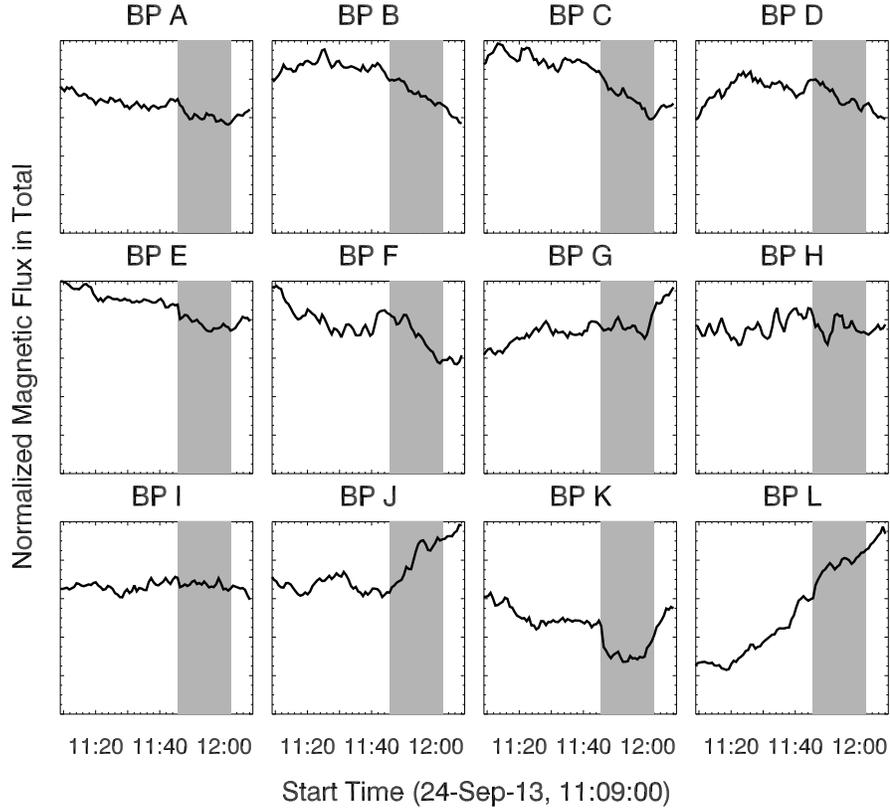}
\caption{\bf The curves show the evolution of the normalized magnetic flux for the selected BPs from 11:09 UT to 12:09 UT. They sum the positive and modulus of the negative magnetic flux. The grey rectangular regions give the time range of the AIA observations and the scanning period of the IRIS in our work.}
\label{fig10}
\end{figure}

\begin{figure}
\centering
\includegraphics[scale=0.5, angle=0]{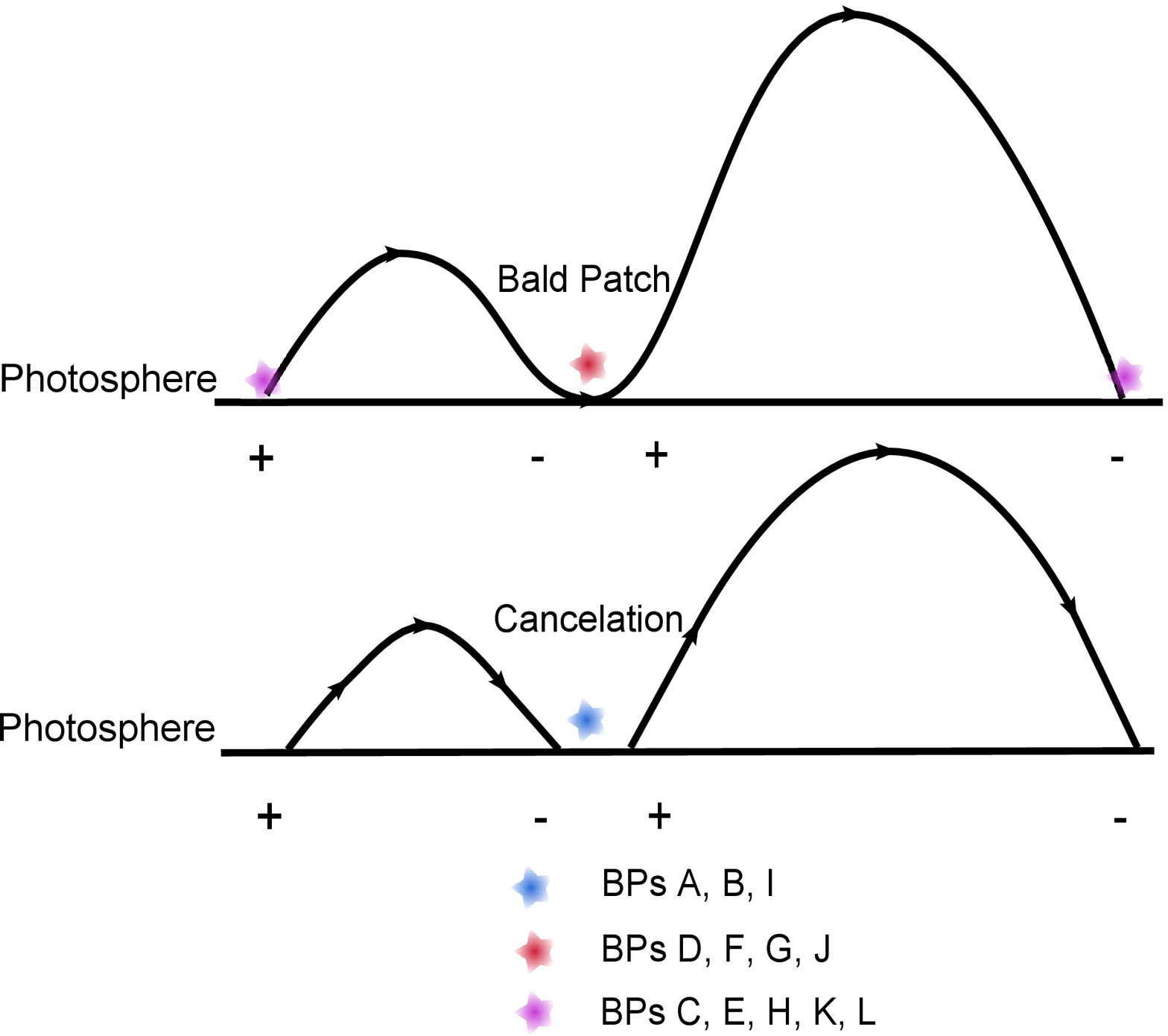}
\caption{\bf Sketch for the magnetic conditions of the BPs A to L.}
\label{fig07}
\end{figure}

\begin{figure}
\centering
\includegraphics[scale=0.7, angle=0, trim = 0.0cm 1.0cm 0.0cm 0.0cm]{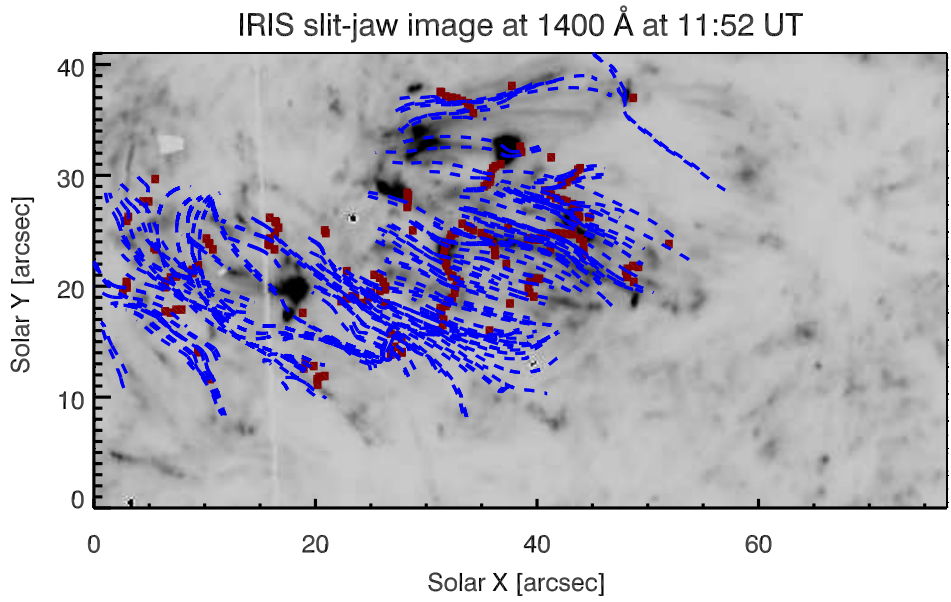}
\caption{Overview of magnetic field line connectivity in AR11850. Red dots mark the locations of bald patches;  Blue  field lines display magnetic field lines passing through the corresponding bald patches obtained by  a forced-field extrapolation.}
\label{fig11}
\end{figure}

\begin{figure}
\centering
\includegraphics[scale=0.6, angle=0, trim = 0.0cm 1.0cm 0.0cm 4.8cm, clip]{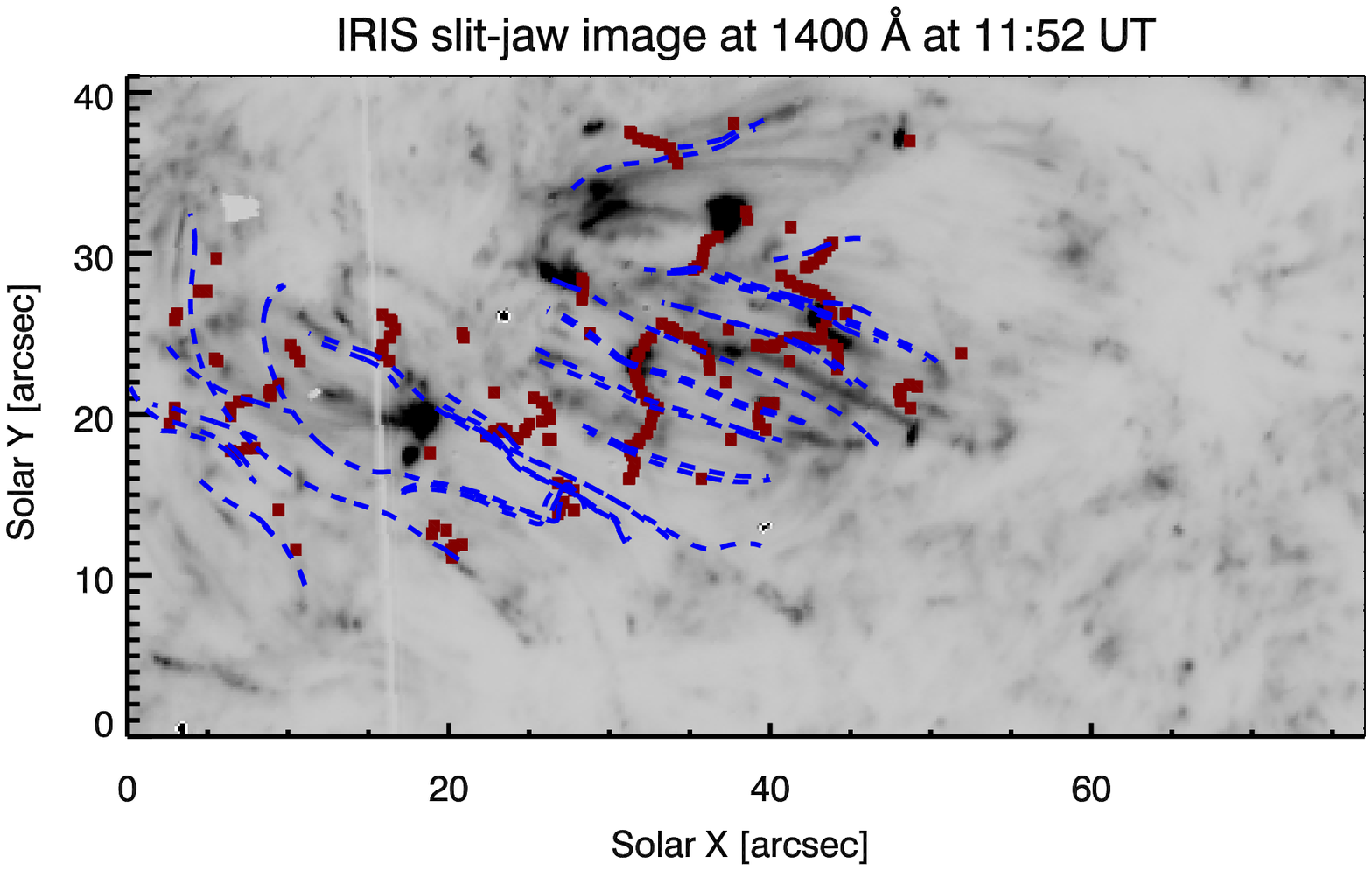}
\includegraphics[scale=0.33, angle=0, trim = 2.0cm 2.0cm 2.0cm 12.8cm, clip]{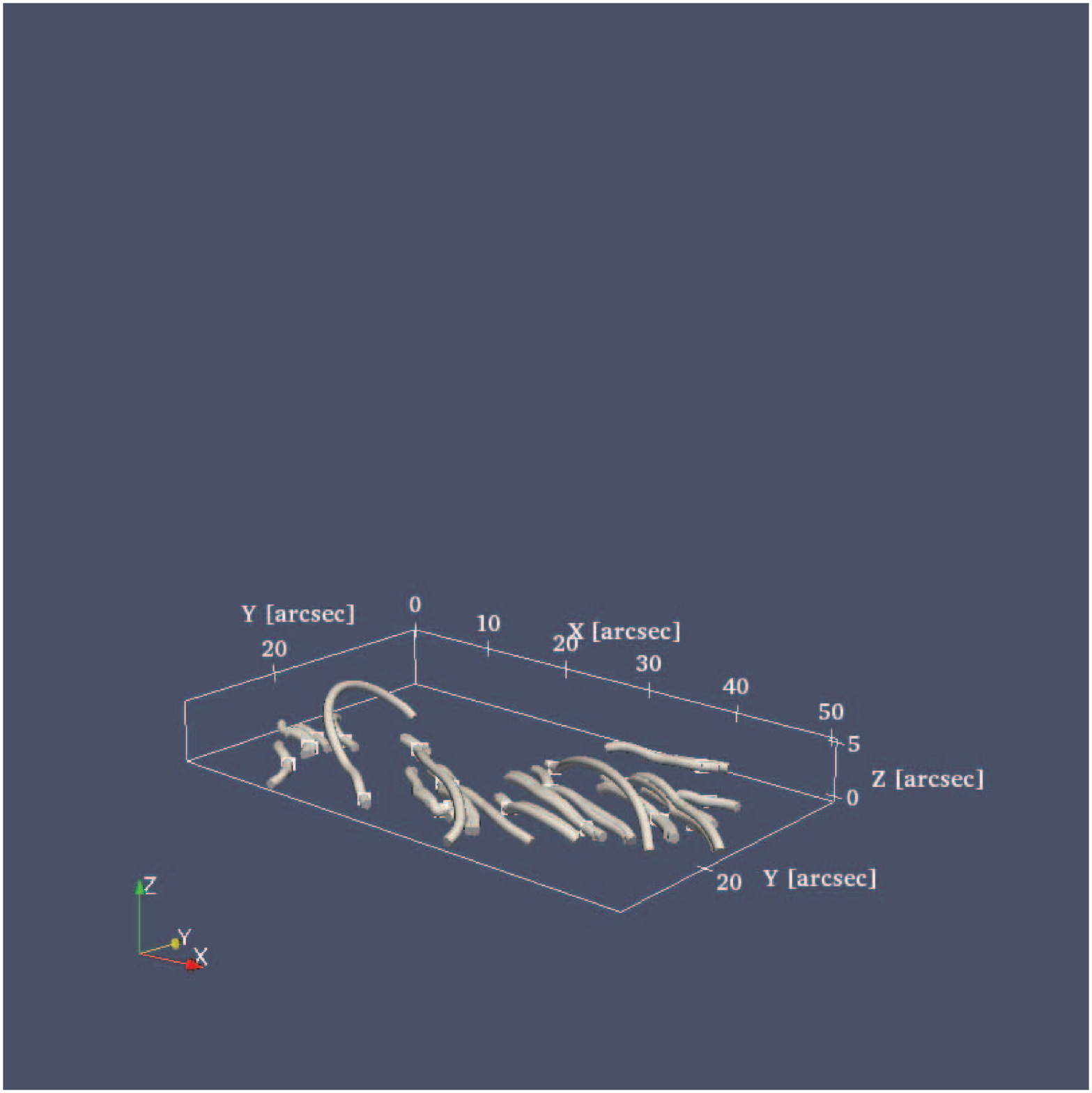}
\includegraphics[scale=0.33, angle=0, trim = 1.5cm 4.0cm 2.5cm 10.8cm, clip]{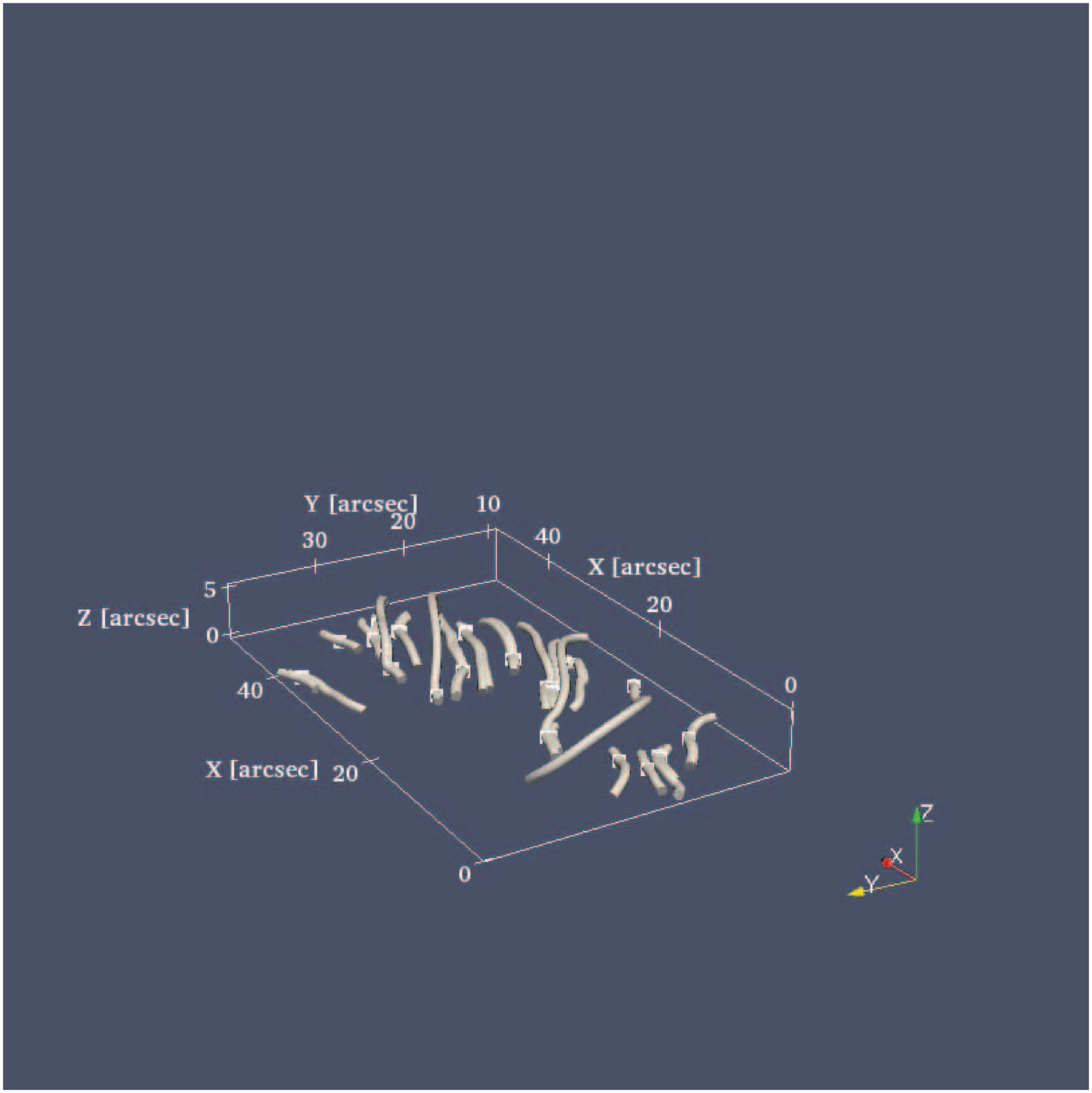}
\caption{Selected magnetic field lines in AR11850. Upper panel: top view; Bottom panels: side views.
The red dots (in the upper panel) and the white dots (in the bottom panels) mark the locations of bald patch.}
\label{fig12}
\end{figure}

We also investigate the temporal evolution of
these BPs in AIA 1600 and 1700 \AA. They always exist during
the period of around 20 minutes. However, the {\bf curve of intensity evolution} shows a
periodic variation {\bf of} several minutes, {\bf which could be probably due to the periodic reconnection.}
According to the scanning time of the IRIS raster,
we determined the moments when the raster scanning through
these BPs (labeled as white rectangular boxes in Figure
\ref{fig03} and \ref{fig04}) and calculated the intensity ratio between
AIA 1600 \AA\ and 1700 \AA\ at these moments. The results suggest
that some BPs, such as A to F, have more contribution
from C IV line, i.e. from higher temperature plasma while the others do not.

As the heating only happens at particular sites, e.g. the BP sites,
it means that the energy release only occurs under special conditions.
For understanding the {\bf non-uniform} heating effect, we have studied the properties
of these BPs in detail to see {\bf how} they are heated and why the
heating effects are different.
{\bf The magnetic configuration and the magnetic flux evolution at the corresponding locations suggest that
BPs A, B and I are consistent with the magnetic cancelation scenario, BPs D, F, G, J appear to
have magnetic bald patch topology, and BPs C, E, H, K, L are located at the footpoints of the bald patch related separatrix layers.
{\bf The magnetic field separatrix layers are volume structures at where the magnetic connectivity changes and current sheet can easily form \citep{1987ApJ...323..358L}.}
Considering the contribution of the C IV line in different BPs, our results indicate
that the bald patch reconnection may have weaker heating effect than the
magnetic cancelation, and the continuous cancelation or reconnection may have stronger heating effect than the temporary one.}

According to the simulation work of \cite{2016arXiv161101746N}, the {\bf level} of the heating effect
{\bf depends on} the local plasma $\beta$, which is
directly proportional with plasma density and {\bf inversely} proportional to magnetic field strength.
Their results show that low $\beta$  magnetic reconnection process is associated with high temperature events and high $\beta$
with low temperature events. In our situation, more materials would be deposited at the bald patch locations
{\bf than} in cancelation regions. Supposing the magnetic field under these two conditions is of the same order,
we suggest that our {\bf observational analyzes support the theoretical results.}

Besides the BP sites, energy release also happens at some
{\bf bright arcade structures.} According to previous studies based on the LFFF or NLFFF extrapolations, or MHD simulation,
these structures are suspected to be related {\bf to} the reconnection
happened at the serpentine field lines that passing through
the bald patches \citep{2007A&A...473..279P,2012SoPh..278...73V,2009ApJ...701.1911P}. With our extrapolation result,
we obtain the corresponding serpentine field lines and suggest that these
lines have good accordance with the bright {\bf arcades} in between the emerging region.
This fact confirms the conclusions of previous
works from  an observational aspect, that is when the new flux emerges, it may
have some difficulty in passing through the photosphere
and becomes horizontal at the sub-surface as the pressure height changes.
It can only continue the emergence until the reconnection
happens at its magnetic dips as well as its separatrix surface under suitable photospheric flows.
{\bf The bright arcades visible in SJI 1400 \AA\ indicate that probable reconnection between the emerging flux and the overlying magnetic flux
occurs in current layers at QSLs, as suggested in the simulation work of \citet{2009ApJ...701.1911P}.}

In summary, we have studied the BPs and {\bf transition region arcades} in an emerging flux region.
{\bf Using} our forced-field extrapolation, we find observational evidence of magnetic reconnection for these structures.
This is the first time for the extrapolation method, which considering the physical condition in the lower atmosphere,
to be used for investigating the {\bf local heating events}.

\acknowledgements
We thank
Dr. T. Wiegelmann in discussing the extrapolation method, P. D\'emoulin for fruitful discussion
on the emerging flux conditions, Dr. H. Tian in understanding the IRIS spectral lines,
{\bf J.W. Li and} Dr. D. Li for helping us with the IRIS data analysis
and Dr. Y. Guo for helping us with the software PARAVIEW.
B.S. would like to thank the team of the ISSI workshop on "Solar UV bursts- a new
insight to magnetic reconnection" Lead by Peter Young for fruitful discussions and ISSI for
its financial support.
J.Z. and H.L. are supported by NSFC under
grant 11273065, and by the Strategic Pioneer Program on
Space Sciences, Chinese Academy of Sciences, under grant
XDA04076101. J.Z. is also supported by NSFC under grants
11503089, 11522328, and 11473070.
IRIS is a NASA small explorer mission developed and operated by
LMSAL with mission operations executed at NASA ARC
and major contributions to downlink communications funded by ESA and the Norwegian Space Center.

\bibliographystyle{apj}
\bibliography{zhao2016_references} 

\begin{thebibliography}{}
\expandafter\ifx\csname natexlab\endcsname\relax\def\natexlab#1{#1}\fi

\bibitem[{{Alipour} \& {Safari}(2015)}]{2015ApJ...807..175A}
{Alipour}, N., \& {Safari}, H. 2015, \apj, 807, 175

\bibitem[{{Aschwanden}(2004)}]{2004psci.book.....A}
{Aschwanden}, M.~J. 2004, {Physics of the Solar Corona. An Introduction}
  (Praxis Publishing Ltd)

\bibitem[{{Aulanier} {et~al.}(1998){Aulanier}, {D{\'e}moulin}, {Schmieder},
  {Fang}, \& {Tang}}]{1998SoPh..183..369A}
{Aulanier}, G., {D{\'e}moulin}, P., {Schmieder}, B., {Fang}, C., \& {Tang},
  Y.~H. 1998, \solphys, 183, 369

\bibitem[{{Berger} {et~al.}(2007){Berger}, {Rouppe van der Voort}, \&
  {L{\"o}fdahl}}]{2007ApJ...661.1272B}
{Berger}, T.~E., {Rouppe van der Voort}, L., \& {L{\"o}fdahl}, M. 2007, \apj,
  661, 1272

\bibitem[{{Brooks} {et~al.}(2016){Brooks}, {Reep}, \&
  {Warren}}]{2016ApJ...826L..18B}
{Brooks}, D.~H., {Reep}, J.~W., \& {Warren}, H.~P. 2016, \apjl, 826, L18

\bibitem[{{Brosius}(2013)}]{2013ApJ...777..135B}
{Brosius}, J.~W. 2013, \apj, 777, 135

\bibitem[{{Brosius} \& {Holman}(2010)}]{2010ApJ...720.1472B}
{Brosius}, J.~W., \& {Holman}, G.~D. 2010, \apj, 720, 1472

\bibitem[{{Chandrashekhar} {et~al.}(2013){Chandrashekhar}, {Krishna Prasad},
  {Banerjee}, {Ravindra}, \& {Seaton}}]{2013SoPh..286..125C}
{Chandrashekhar}, K., {Krishna Prasad}, S., {Banerjee}, D., {Ravindra}, B., \&
  {Seaton}, D.~B. 2013, \solphys, 286, 125

\bibitem[{{Chen} \& {Ding}(2010)}]{2010ApJ...724..640C}
{Chen}, F., \& {Ding}, M.~D. 2010, \apj, 724, 640

\bibitem[{{Cheng} {et~al.}(2015){Cheng}, {Ding}, \&
  {Fang}}]{2015ApJ...804...82C}
{Cheng}, X., {Ding}, M.~D., \& {Fang}, C. 2015, \apj, 804, 82

\bibitem[{{Chesny} {et~al.}(2015){Chesny}, {Oluseyi}, {Orange}, \&
  {Champey}}]{2015ApJ...814..124C}
{Chesny}, D.~L., {Oluseyi}, H.~M., {Orange}, N.~B., \& {Champey}, P.~R. 2015,
  \apj, 814, 124

\bibitem[{{De Pontieu} {et~al.}(2014{\natexlab{a}}){De Pontieu}, {Rouppe van
  der Voort}, {McIntosh}, {Pereira}, {Carlsson}, {Hansteen}, {Skogsrud},
  {Lemen}, {Title}, {Boerner}, {Hurlburt}, {Tarbell}, {Wuelser}, {De Luca},
  {Golub}, {McKillop}, {Reeves}, {Saar}, {Testa}, {Tian}, {Kankelborg},
  {Jaeggli}, {Kleint}, \& {Martinez-Sykora}}]{2014Sci...346D.315D}
{De Pontieu}, B., {Rouppe van der Voort}, L., {McIntosh}, S.~W., {et~al.}
  2014{\natexlab{a}}, Science, 346, 1255732

\bibitem[{{De Pontieu} {et~al.}(2014{\natexlab{b}}){De Pontieu}, {Title},
  {Lemen}, {Kushner}, {Akin}, {Allard}, {Berger}, {Boerner}, {Cheung}, {Chou},
  {Drake}, {Duncan}, {Freeland}, {Heyman}, {Hoffman}, {Hurlburt}, {Lindgren},
  {Mathur}, {Rehse}, {Sabolish}, {Seguin}, {Schrijver}, {Tarbell},
  {W{\"u}lser}, {Wolfson}, {Yanari}, {Mudge}, {Nguyen-Phuc}, {Timmons}, {van
  Bezooijen}, {Weingrod}, {Brookner}, {Butcher}, {Dougherty}, {Eder},
  {Knagenhjelm}, {Larsen}, {Mansir}, {Phan}, {Boyle}, {Cheimets}, {DeLuca},
  {Golub}, {Gates}, {Hertz}, {McKillop}, {Park}, {Perry}, {Podgorski},
  {Reeves}, {Saar}, {Testa}, {Tian}, {Weber}, {Dunn}, {Eccles}, {Jaeggli},
  {Kankelborg}, {Mashburn}, {Pust}, {Springer}, {Carvalho}, {Kleint}, {Marmie},
  {Mazmanian}, {Pereira}, {Sawyer}, {Strong}, {Worden}, {Carlsson}, {Hansteen},
  {Leenaarts}, {Wiesmann}, {Aloise}, {Chu}, {Bush}, {Scherrer}, {Brekke},
  {Martinez-Sykora}, {Lites}, {McIntosh}, {Uitenbroek}, {Okamoto}, {Gummin},
  {Auker}, {Jerram}, {Pool}, \& {Waltham}}]{2014SoPh..289.2733D}
{De Pontieu}, B., {Title}, A.~M., {Lemen}, J.~R., {et~al.} 2014{\natexlab{b}},
  \solphys, 289, 2733

\bibitem[{{Doyle} {et~al.}(2006){Doyle}, {Popescu}, \&
  {Taroyan}}]{2006A&A...446..327D}
{Doyle}, J.~G., {Popescu}, M.~D., \& {Taroyan}, Y. 2006, \aap, 446, 327

\bibitem[{{Fan}(2001{\natexlab{a}})}]{2001ApJ...546..509F}
{Fan}, Y. 2001{\natexlab{a}}, \apj, 546, 509

\bibitem[{{Fan}(2001{\natexlab{b}})}]{2001ApJ...554L.111F}
---. 2001{\natexlab{b}}, \apjl, 554, L111

\bibitem[{{Feng} {et~al.}(2007){Feng}, {Inhester}, {Solanki}, {Wiegelmann},
  {Podlipnik}, {Howard}, \& {Wuelser}}]{2007ApJ...671L.205F}
{Feng}, L., {Inhester}, B., {Solanki}, S.~K., {et~al.} 2007, \apjl, 671, L205

\bibitem[{{Feng} {et~al.}(2013){Feng}, {Deng}, {Yang}, \&
  {Ji}}]{2013Ap&SS.348...17F}
{Feng}, S., {Deng}, L., {Yang}, Y., \& {Ji}, K. 2013, \apss, 348, 17

\bibitem[{{Goode} {et~al.}(2003){Goode}, {Denker}, {Didkovsky}, {Kuhn}, \&
  {Wang}}]{2003JKAS...36S.125G}
{Goode}, P.~R., {Denker}, C.~J., {Didkovsky}, L.~I., {Kuhn}, J.~R., \& {Wang},
  H. 2003, Journal of Korean Astronomical Society, 36, S125

\bibitem[{{Grubecka} {et~al.}(2016){Grubecka}, {Schmieder}, {Berlicki},
  {Heinzel}, {Dalmasse}, \& {Mein}}]{2016A&A...593A..32G}
{Grubecka}, M., {Schmieder}, B., {Berlicki}, A., {et~al.} 2016, \aap, 593, A32

\bibitem[{{Guo} {et~al.}(2013){Guo}, {D{\'e}moulin}, {Schmieder}, {Ding},
  {Vargas Dom{\'{\i}}nguez}, \& {Liu}}]{2013A&A...555A..19G}
{Guo}, Y., {D{\'e}moulin}, P., {Schmieder}, B., {et~al.} 2013, \aap, 555, A19

\bibitem[{{Hansteen} {et~al.}(2014){Hansteen}, {De Pontieu}, {Carlsson},
  {Lemen}, {Title}, {Boerner}, {Hurlburt}, {Tarbell}, {Wuelser}, {Pereira}, {De
  Luca}, {Golub}, {McKillop}, {Reeves}, {Saar}, {Testa}, {Tian}, {Kankelborg},
  {Jaeggli}, {Kleint}, \& {Mart{\'{\i}}nez-Sykora}}]{2014Sci...346E.315H}
{Hansteen}, V., {De Pontieu}, B., {Carlsson}, M., {et~al.} 2014, Science, 346,
  1255757

\bibitem[{{Harrison}(1997)}]{1997SoPh..175..467H}
{Harrison}, R.~A. 1997, \solphys, 175, 467

\bibitem[{{Heinzel}(1995)}]{1995A&A...299..563H}
{Heinzel}, P. 1995, \aap, 299, 563

\bibitem[{{Hong} {et~al.}(2016){Hong}, {Ding}, {Li}, {Yang}, {Cheng}, {Chen},
  {Fang}, \& {Cao}}]{2016ApJ...820L..17H}
{Hong}, J., {Ding}, M.~D., {Li}, Y., {et~al.} 2016, \apjl, 820, L17

\bibitem[{{Huang} {et~al.}(2014){Huang}, {Madjarska}, {Xia}, {Doyle},
  {Galsgaard}, \& {Fu}}]{2014ApJ...797...88H}
{Huang}, Z., {Madjarska}, M.~S., {Xia}, L., {et~al.} 2014, \apj, 797, 88

\bibitem[{{Innes} {et~al.}(2016){Innes}, {Bu{\v c}{\'{\i}}k}, {Guo}, \&
  {Nitta}}]{2016AN....337.1024I}
{Innes}, D.~E., {Bu{\v c}{\'{\i}}k}, R., {Guo}, L.-J., \& {Nitta}, N. 2016,
  Astronomische Nachrichten, 337, 1024

\bibitem[{{Jafarzadeh} {et~al.}(2014){Jafarzadeh}, {Cameron}, {Solanki},
  {Pietarila}, {Feller}, {Lagg}, \& {Gandorfer}}]{2014A&A...563A.101J}
{Jafarzadeh}, S., {Cameron}, R.~H., {Solanki}, S.~K., {et~al.} 2014, \aap, 563,
  A101

\bibitem[{{Janvier} {et~al.}(2015){Janvier}, {Aulanier}, \&
  {D{\'e}moulin}}]{2015SoPh..290.3425J}
{Janvier}, M., {Aulanier}, G., \& {D{\'e}moulin}, P. 2015, \solphys, 290, 3425

\bibitem[{{Ji} {et~al.}(2016{\natexlab{a}}){Ji}, {Jiang}, {Feng}, {Yang},
  {Deng}, \& {Wang}}]{2016SoPh..291..357J}
{Ji}, K., {Jiang}, X., {Feng}, S., {et~al.} 2016{\natexlab{a}}, \solphys, 291,
  357

\bibitem[{{Ji} {et~al.}(2016{\natexlab{b}}){Ji}, {Xiong}, {Xiang}, {Feng},
  {Deng}, {Wang}, \& {Yang}}]{2016RAA....16e...9J}
{Ji}, K.-F., {Xiong}, J.-P., {Xiang}, Y.-Y., {et~al.} 2016{\natexlab{b}},
  Research in Astronomy and Astrophysics, 16, 009

\bibitem[{{Jiang} {et~al.}(2015){Jiang}, {Zhang}, \&
  {Yang}}]{2015PASJ...67...40J}
{Jiang}, F., {Zhang}, J., \& {Yang}, S. 2015, \pasj, 67, 40

\bibitem[{{Jiang} {et~al.}(2010){Jiang}, {Fang}, \&
  {Chen}}]{2010ApJ...710.1387J}
{Jiang}, R.~L., {Fang}, C., \& {Chen}, P.~F. 2010, \apj, 710, 1387

\bibitem[{{Jiang} {et~al.}(2012){Jiang}, {Fang}, \&
  {Chen}}]{2012ApJ...751..152J}
{Jiang}, R.-L., {Fang}, C., \& {Chen}, P.-F. 2012, \apj, 751, 152

\bibitem[{{Kamio} {et~al.}(2011){Kamio}, {Curdt}, {Teriaca}, \&
  {Innes}}]{2011A&A...529A..21K}
{Kamio}, S., {Curdt}, W., {Teriaca}, L., \& {Innes}, D.~E. 2011, \aap, 529, A21

\bibitem[{{Kim} {et~al.}(2015){Kim}, {Yurchyshyn}, {Bong}, {Cho}, {Cho}, {Lee},
  {Lim}, {Park}, {Yang}, {Ahn}, {Goode}, \& {Jang}}]{2015ApJ...810...38K}
{Kim}, Y.-H., {Yurchyshyn}, V., {Bong}, S.-C., {et~al.} 2015, \apj, 810, 38

\bibitem[{{Leiko} \& {Kondrashova}(2015)}]{2015AdSpR..55..886L}
{Leiko}, U.~M., \& {Kondrashova}, N.~N. 2015, Advances in Space Research, 55,
  886

\bibitem[{{Lemen} {et~al.}(2012){Lemen}, {Title}, {Akin}, {Boerner}, {Chou},
  {Drake}, {Duncan}, {Edwards}, {Friedlaender}, {Heyman}, {Hurlburt}, {Katz},
  {Kushner}, {Levay}, {Lindgren}, {Mathur}, {McFeaters}, {Mitchell}, {Rehse},
  {Schrijver}, {Springer}, {Stern}, {Tarbell}, {Wuelser}, {Wolfson}, {Yanari},
  {Bookbinder}, {Cheimets}, {Caldwell}, {Deluca}, {Gates}, {Golub}, {Park},
  {Podgorski}, {Bush}, {Scherrer}, {Gummin}, {Smith}, {Auker}, {Jerram},
  {Pool}, {Soufli}, {Windt}, {Beardsley}, {Clapp}, {Lang}, \&
  {Waltham}}]{2012SoPh..275...17L}
{Lemen}, J.~R., {Title}, A.~M., {Akin}, D.~J., {et~al.} 2012, \solphys, 275, 17

\bibitem[{{Li} \& {Ning}(2012)}]{2012Ap&SS.341..215L}
{Li}, D., \& {Ning}, Z. 2012, \apss, 341, 215

\bibitem[{{Li} {et~al.}(2013){Li}, {Ning}, \& {Wang}}]{2013NewA...23...19L}
{Li}, D., {Ning}, Z.~J., \& {Wang}, J.~F. 2013, \na, 23, 19

\bibitem[{{Li} {et~al.}(2007){Li}, {Sakurai}, {Ichimito}, {Suematsu},
  {Tsuneta}, {Katsukawa}, {Shimizu}, {Shine}, {Tarbell}, {Title}, {Lites},
  {Kubo}, {Nagata}, {Kotoku}, {Shibasaki}, {Saar}, \&
  {Bobra}}]{2007PASJ...59S.643L}
{Li}, H., {Sakurai}, T., {Ichimito}, K., {et~al.} 2007, \pasj, 59, S643

\bibitem[{{Li} \& {Zhang}(2016)}]{2016A&A...589A.114L}
{Li}, T., \& {Zhang}, J. 2016, \aap, 589, A114

\bibitem[{{Low}(1987)}]{1987ApJ...323..358L}
{Low}, B.~C. 1987, \apj, 323, 358

\bibitem[{{Madjarska} {et~al.}(2003){Madjarska}, {Doyle}, {Teriaca}, \&
  {Banerjee}}]{2003A&A...398..775M}
{Madjarska}, M.~S., {Doyle}, J.~G., {Teriaca}, L., \& {Banerjee}, D. 2003,
  \aap, 398, 775

\bibitem[{{Mart{\'{\i}}nez-Sykora} {et~al.}(2015){Mart{\'{\i}}nez-Sykora},
  {Rouppe van der Voort}, {Carlsson}, {De Pontieu}, {Pereira}, {Boerner},
  {Hurlburt}, {Kleint}, {Lemen}, {Tarbell}, {Title}, {Wuelser}, {Hansteen},
  {Golub}, {McKillop}, {Reeves}, {Saar}, {Testa}, {Tian}, {Jaeggli}, \&
  {Kankelborg}}]{2015ApJ...803...44M}
{Mart{\'{\i}}nez-Sykora}, J., {Rouppe van der Voort}, L., {Carlsson}, M.,
  {et~al.} 2015, \apj, 803, 44

\bibitem[{{Milligan}(2015)}]{2015SoPh..290.3399M}
{Milligan}, R.~O. 2015, \solphys, 290, 3399

\bibitem[{{Mou} {et~al.}(2016){Mou}, {Huang}, {Xia}, {Madjarska}, {Li}, {Fu},
  {Jiao}, \& {Hou}}]{2016ApJ...818....9M}
{Mou}, C., {Huang}, Z., {Xia}, L., {et~al.} 2016, \apj, 818, 9

\bibitem[{{Ni} {et~al.}(2015{\natexlab{a}}){Ni}, {Kliem}, {Lin}, \&
  {Wu}}]{2015ApJ...799...79N}
{Ni}, L., {Kliem}, B., {Lin}, J., \& {Wu}, N. 2015{\natexlab{a}}, \apj, 799, 79

\bibitem[{{Ni} {et~al.}(2015{\natexlab{b}}){Ni}, {Lin}, {Mei}, \&
  {Li}}]{2015ApJ...812...92N}
{Ni}, L., {Lin}, J., {Mei}, Z., \& {Li}, Y. 2015{\natexlab{b}}, \apj, 812, 92

\bibitem[{{Ni} {et~al.}(2016){Ni}, {Lin}, {Roussev}, \&
  {Schmieder}}]{2016arXiv161101746N}
{Ni}, L., {Lin}, J., {Roussev}, I.~I., \& {Schmieder}, B. 2016, ArXiv e-prints,
  arXiv:1611.01746

\bibitem[{{Ning} \& {Guo}(2014)}]{2014ApJ...794...79N}
{Ning}, Z., \& {Guo}, Y. 2014, \apj, 794, 79

\bibitem[{{Pariat} {et~al.}(2010){Pariat}, {Antiochos}, \&
  {DeVore}}]{2010ApJ...714.1762P}
{Pariat}, E., {Antiochos}, S.~K., \& {DeVore}, C.~R. 2010, \apj, 714, 1762

\bibitem[{{Pariat} {et~al.}(2004){Pariat}, {Aulanier}, {Schmieder},
  {Georgoulis}, {Rust}, \& {Bernasconi}}]{2004ApJ...614.1099P}
{Pariat}, E., {Aulanier}, G., {Schmieder}, B., {et~al.} 2004, \apj, 614, 1099

\bibitem[{{Pariat} {et~al.}(2006){Pariat}, {Aulanier}, {Schmieder},
  {Georgoulis}, {Rust}, \& {Bernasconi}}]{2006AdSpR..38..902P}
---. 2006, Advances in Space Research, 38, 902

\bibitem[{{Pariat} {et~al.}(2009){Pariat}, {Masson}, \&
  {Aulanier}}]{2009ApJ...701.1911P}
{Pariat}, E., {Masson}, S., \& {Aulanier}, G. 2009, \apj, 701, 1911

\bibitem[{{Pariat} {et~al.}(2007){Pariat}, {Schmieder}, {Berlicki}, {Deng},
  {Mein}, {L{\'o}pez Ariste}, \& {Wang}}]{2007A&A...473..279P}
{Pariat}, E., {Schmieder}, B., {Berlicki}, A., {et~al.} 2007, \aap, 473, 279

\bibitem[{{Park} {et~al.}(2016){Park}, {Tsiropoula}, {Kontogiannis},
  {Tziotziou}, {Scullion}, \& {Doyle}}]{2016A&A...586A..25P}
{Park}, S.-H., {Tsiropoula}, G., {Kontogiannis}, I., {et~al.} 2016, \aap, 586,
  A25

\bibitem[{{Parker}(1981)}]{1981GApFD..18..332P}
{Parker}, E.~N. 1981, Geophysical and Astrophysical Fluid Dynamics, 18, 332

\bibitem[{{Parnell} {et~al.}(2002){Parnell}, {Bewsher}, \&
  {Harrison}}]{2002SoPh..206..249P}
{Parnell}, C.~E., {Bewsher}, D., \& {Harrison}, R.~A. 2002, \solphys, 206, 249

\bibitem[{{Peter} {et~al.}(2014){Peter}, {Tian}, {Curdt}, {Schmit}, {Innes},
  {De Pontieu}, {Lemen}, {Title}, {Boerner}, {Hurlburt}, {Tarbell}, {Wuelser},
  {Mart{\'{\i}}nez-Sykora}, {Kleint}, {Golub}, {McKillop}, {Reeves}, {Saar},
  {Testa}, {Kankelborg}, {Jaeggli}, {Carlsson}, \&
  {Hansteen}}]{2014Sci...346C.315P}
{Peter}, H., {Tian}, H., {Curdt}, W., {et~al.} 2014, Science, 346, 1255726

\bibitem[{{Phillips}(1991)}]{1991VA.....34..353P}
{Phillips}, K.~J.~H. 1991, Vistas in Astronomy, 34, 353

\bibitem[{{Potts} {et~al.}(2007){Potts}, {Khan}, \&
  {Diver}}]{2007SoPh..245...55P}
{Potts}, H.~E., {Khan}, J.~I., \& {Diver}, D.~A. 2007, \solphys, 245, 55

\bibitem[{{Roumeliotis}(1996)}]{1996ApJ...473.1095R}
{Roumeliotis}, G. 1996, \apj, 473, 1095

\bibitem[{{Rouppe van der Voort} {et~al.}(2015){Rouppe van der Voort}, {De
  Pontieu}, {Pereira}, {Carlsson}, \& {Hansteen}}]{2015ApJ...799L...3R}
{Rouppe van der Voort}, L., {De Pontieu}, B., {Pereira}, T.~M.~D., {Carlsson},
  M., \& {Hansteen}, V. 2015, \apjl, 799, L3

\bibitem[{{Sakurai}(1982)}]{1982SoPh...76..301S}
{Sakurai}, T. 1982, \solphys, 76, 301

\bibitem[{{Samanta} {et~al.}(2015){Samanta}, {Banerjee}, \&
  {Tian}}]{2015ApJ...806..172S}
{Samanta}, T., {Banerjee}, D., \& {Tian}, H. 2015, \apj, 806, 172

\bibitem[{{Santos} \& {B{\"u}chner}(2007)}]{2007ASTRA...3...29S}
{Santos}, J.~C., \& {B{\"u}chner}, J. 2007, Astrophysics and Space Sciences
  Transactions, 3, 29

\bibitem[{{Scherrer} {et~al.}(2012){Scherrer}, {Schou}, {Bush}, {Kosovichev},
  {Bogart}, {Hoeksema}, {Liu}, {Duvall}, {Zhao}, {Title}, {Schrijver},
  {Tarbell}, \& {Tomczyk}}]{2012SoPh..275..207S}
{Scherrer}, P.~H., {Schou}, J., {Bush}, R.~I., {et~al.} 2012, \solphys, 275,
  207

\bibitem[{{Schmieder}(1997)}]{1997LNP...489..139S}
{Schmieder}, B. 1997, in Lecture Notes in Physics, Berlin Springer Verlag, Vol.
  489, European Meeting on Solar Physics, ed. G.~M. {Simnett}, C.~E.
  {Alissandrakis}, \& L.~{Vlahos}, 139

\bibitem[{{Schmieder} {et~al.}(2014){Schmieder}, {Archontis}, \&
  {Pariat}}]{2014SSRv..186..227S}
{Schmieder}, B., {Archontis}, V., \& {Pariat}, E. 2014, \ssr, 186, 227

\bibitem[{{Schmieder} \& {Aulanier}(2003)}]{2003AdSpR..32.1875S}
{Schmieder}, B., \& {Aulanier}, G. 2003, Advances in Space Research, 32, 1875

\bibitem[{{Schmieder} {et~al.}(2004){Schmieder}, {Rust}, {Georgoulis},
  {D{\'e}moulin}, \& {Bernasconi}}]{2004ApJ...601..530S}
{Schmieder}, B., {Rust}, D.~M., {Georgoulis}, M.~K., {D{\'e}moulin}, P., \&
  {Bernasconi}, P.~N. 2004, \apj, 601, 530

\bibitem[{{Schmieder} {et~al.}(1995){Schmieder}, {Shibata}, {van
  Driel-Gesztelyi}, \& {Freeland}}]{1995SoPh..156..245S}
{Schmieder}, B., {Shibata}, K., {van Driel-Gesztelyi}, L., \& {Freeland}, S.
  1995, \solphys, 156, 245

\bibitem[{{Schrijver}(2009)}]{2009AdSpR..43..739S}
{Schrijver}, C.~J. 2009, Advances in Space Research, 43, 739

\bibitem[{{Shimizu}(1995)}]{1995PASJ...47..251S}
{Shimizu}, T. 1995, \pasj, 47, 251

\bibitem[{{Skogsrud} {et~al.}(2016){Skogsrud}, {Rouppe van der Voort}, \& {De
  Pontieu}}]{2016ApJ...817..124S}
{Skogsrud}, H., {Rouppe van der Voort}, L., \& {De Pontieu}, B. 2016, \apj,
  817, 124

\bibitem[{{Tian} {et~al.}(2010){Tian}, {Potts}, {Marsch}, {Attie}, \&
  {He}}]{2010A&A...519A..58T}
{Tian}, H., {Potts}, H.~E., {Marsch}, E., {Attie}, R., \& {He}, J.-S. 2010,
  \aap, 519, A58

\bibitem[{{Tian} {et~al.}(2016){Tian}, {Xu}, {He}, \&
  {Madsen}}]{2016ApJ...824...96T}
{Tian}, H., {Xu}, Z., {He}, J., \& {Madsen}, C. 2016, \apj, 824, 96

\bibitem[{{Tian} {et~al.}(2014){Tian}, {DeLuca}, {Cranmer}, {De Pontieu},
  {Peter}, {Mart{\'{\i}}nez-Sykora}, {Golub}, {McKillop}, {Reeves}, {Miralles},
  {McCauley}, {Saar}, {Testa}, {Weber}, {Murphy}, {Lemen}, {Title}, {Boerner},
  {Hurlburt}, {Tarbell}, {Wuelser}, {Kleint}, {Kankelborg}, {Jaeggli},
  {Carlsson}, {Hansteen}, \& {McIntosh}}]{2014Sci...346A.315T}
{Tian}, H., {DeLuca}, E.~E., {Cranmer}, S.~R., {et~al.} 2014, Science, 346,
  1255711

\bibitem[{{Titov} {et~al.}(1993){Titov}, {Priest}, \&
  {Demoulin}}]{1993A&A...276..564T}
{Titov}, V.~S., {Priest}, E.~R., \& {Demoulin}, P. 1993, \aap, 276, 564

\bibitem[{{Ugarte-Urra} {et~al.}(2004){Ugarte-Urra}, {Doyle}, {Madjarska}, \&
  {O'Shea}}]{2004A&A...418..313U}
{Ugarte-Urra}, I., {Doyle}, J.~G., {Madjarska}, M.~S., \& {O'Shea}, E. 2004,
  \aap, 418, 313

\bibitem[{{Valori} {et~al.}(2012){Valori}, {Green}, {D{\'e}moulin}, {Vargas
  Dom{\'{\i}}nguez}, {van Driel-Gesztelyi}, {Wallace}, {Baker}, \&
  {Fuhrmann}}]{2012SoPh..278...73V}
{Valori}, G., {Green}, L.~M., {D{\'e}moulin}, P., {et~al.} 2012, \solphys, 278,
  73

\bibitem[{{Vissers} {et~al.}(2015){Vissers}, {Rouppe van der Voort}, {Rutten},
  {Carlsson}, \& {De Pontieu}}]{2015ApJ...812...11V}
{Vissers}, G.~J.~M., {Rouppe van der Voort}, L.~H.~M., {Rutten}, R.~J.,
  {Carlsson}, M., \& {De Pontieu}, B. 2015, \apj, 812, 11

\bibitem[{{Wheatland}(2004)}]{2004SoPh..222..247W}
{Wheatland}, M.~S. 2004, \solphys, 222, 247

\bibitem[{{Wiegelmann} \& {Inhester}(2010)}]{2010A&A...516A.107W}
{Wiegelmann}, T., \& {Inhester}, B. 2010, \aap, 516, A107

\bibitem[{{Yang} {et~al.}(2011){Yang}, {Zhang}, {Li}, \&
  {Liu}}]{2011ApJ...732L...7Y}
{Yang}, S., {Zhang}, J., {Li}, T., \& {Liu}, Y. 2011, \apjl, 732, L7

\bibitem[{{Yang} {et~al.}(2015){Yang}, {Ji}, {Feng}, {Deng}, {Wang}, \&
  {Lin}}]{2015ApJ...810...88Y}
{Yang}, Y., {Ji}, K., {Feng}, S., {et~al.} 2015, \apj, 810, 88

\bibitem[{{Yang} {et~al.}(2016){Yang}, {Li}, {Ji}, {Feng}, {Deng}, {Wang}, \&
  {Lin}}]{2016SoPh..291.1089Y}
{Yang}, Y., {Li}, Q., {Ji}, K., {et~al.} 2016, \solphys, 291, 1089

\bibitem[{{Zhang} {et~al.}(2001){Zhang}, {Kundu}, \&
  {White}}]{2001SoPh..198..347Z}
{Zhang}, J., {Kundu}, M.~R., \& {White}, S.~M. 2001, \solphys, 198, 347

\bibitem[{{Zhang} {et~al.}(2014){Zhang}, {Chen}, {Ding}, \&
  {Ji}}]{2014A&A...568A..30Z}
{Zhang}, Q.~M., {Chen}, P.~F., {Ding}, M.~D., \& {Ji}, H.~S. 2014, \aap, 568,
  A30

\bibitem[{{Zhang} {et~al.}(2012){Zhang}, {Chen}, {Guo}, {Fang}, \&
  {Ding}}]{2012ApJ...746...19Z}
{Zhang}, Q.~M., {Chen}, P.~F., {Guo}, Y., {Fang}, C., \& {Ding}, M.~D. 2012,
  \apj, 746, 19

\bibitem[{{Zhao} \& {Li}(2012)}]{2012RAA....12.1681Z}
{Zhao}, J., \& {Li}, H. 2012, Research in Astronomy and Astrophysics, 12, 1681

\bibitem[{{Zhu} {et~al.}(2013){Zhu}, {Wang}, {Du}, \&
  {Fan}}]{2013ApJ...768..119Z}
{Zhu}, X.~S., {Wang}, H.~N., {Du}, Z.~L., \& {Fan}, Y.~L. 2013, \apj, 768, 119

\bibitem[{{Zhu} {et~al.}(2016){Zhu}, {Wang}, {Du}, \&
  {He}}]{2016ApJ...826...51X}
{Zhu}, X.~S., {Wang}, H.~N., {Du}, Z.~L., \& {He}, H. 2016, \apj, 826, 51

\end{thebibliography}

\begin{table}[!htbp]
\centering
\caption[]{Intensity ratio of 1600 \AA\ to 1700 \AA\ of the selected bright points A-I}
\label{Table1}
\begin{tabular}{ccccccc}
\hline\noalign{\smallskip}
Bright Point  & Ratio & Bright Point &  Ratio & Bright Point & Ratio\\
\hline\noalign{\smallskip}
A(Bomb 3) & 1.29 & E& 1.25 & I(Bomb 2)& 0.95 \\
B(Bomb 4) & 1.52 & F& 1.19 & J& 1.04 \\
C & 1.13 & G& 1.04 & K& 1.02 \\
D(Bomb 1) & 1.16 & H& 1.03 & L& 1.06 \\
\noalign{\smallskip}\hline
\end{tabular}
\end{table}

\end{document}